\documentclass[aps,prd,twocolumn,groupedaddress, showkeys, showpacs]{revtex4}
\usepackage{graphicx,subfigure,enumerate,pstricks,latexsym,longtable}
\usepackage[english]{babel}

\newcommand{\Schw}{Schwarzschild}

\newcommand{\beq}{\begin{equation}}
\newcommand{\eeq}{\end{equation}}
\newcommand{\bea}{\begin{eqnarray}}
\newcommand{\eea}{\end{eqnarray}}

\newcommand{\aaa}{{\diamond}}

\newcommand{\ce}{{\cal{E}}} \newcommand{\cl}{{\cal{L}}} \newcommand{\cb}{{\cal{B}}}

\def\p{p}  \def\cp{\pi}
\def\x{x} 

\def\lin{ --- }

\def\af{\zeta}

\def\rr{r}
\def\tt{\theta}

\def\dr{\delta r}
\def\dt{\delta \theta}

\providecommand{\dif}{\mathrm{d}} \def\d{\dif}

\def\mir{\mathrm{r}}
\def\mit{\mathrm{\theta}}
\def\mip{\mathrm{\phi}}
\def\mil{\mathrm{L}}

\def\aprx{\sim}

\begin{document}

\title{Quasi-harmonic oscillatory motion of charged particles around a Schwarzschild black hole immersed in an uniform magnetic field}

\author{
Martin Kolo\v{s}, Zden\v{e}k Stuchl{\'i}k and Arman Tursunov
}
\affiliation{Institute of Physics, Faculty of Philosophy and Science, Silesian University in Opava, \\
Bezru{\v c}ovo n{\'a}m.13, CZ-74601 Opava, Czech Republic}

\begin{abstract}
In order to test the role of large-scale magnetic fields in quasiperiodic oscillation phenomena observed in microquasars, we study oscillatory motion of charged particles in vicinity of a Schwarzschild black hole immersed into an external asymptotically uniform magnetic field. We determine the fundamental frequencies of small harmonic oscillations of charged test particles around stable circular orbits in the equatorial plane of a magnetized black hole, and discuss the radial profiles of frequencies of the radial and latitudinal harmonic oscillations in dependence on the mass of the black hole and the strength of the magnetic field. We demonstrate that assuming relevance of resonant phenomena of the radial and latitudinal oscillations of charged particles at their frequency ratio $3:2$, the oscillatory frequencies of charged particles can be well related to the frequencies of the twin high-frequency quasi-periodic oscillations observed in the microquasars GRS 1915+105, XTE 1550-564 and GRO 1655-40.
\end{abstract}

\date{\today}

 

\maketitle

\section{Introduction}

Investigation of the high-frequency quasi-periodic oscillations (HF QPOs) observed in many black hole or neutron star low-mass X-ray binaries (LMXB) can open up prospects of understanding the phenomena occurring in the presence of strong gravity. Thanks to the possibilities to achieve high precision in the measurements of the frequencies of observed oscillations one can get some useful information about the central object and the electromagnetic fields in its vicinity from the analysis of obtained frequencies. 

In the LMXB systems containing a black hole or a neutron star, the~HF~QPOs are sometimes observed in pairs of the~upper and lower frequencies ($\nu_{\mathrm{U}}$, $\nu_{\mathrm{L}}$) of twin peaks in the~Fourier power spectra. In so called microquasars, i.e., LMXB systems containing a black hole, the twin HF QPOs occurs at fixed frequencies that are usually at the exact $3:2$ ratio \cite{McC-etal:2011:CLAQG:}. The observed high frequencies are close to the~orbital frequency of the~marginally stable circular orbit representing the~inner edge of the Keplerian discs orbiting black holes, therefore, the~strong gravity effects have to be relevant in explaining HF~QPOs \cite{Tor-etal:2005:ASTRA:}. Usually, the Keplerian orbital and epicyclic (radial and latitudinal) frequencies of the circular geodesics of the Kerr geometry \cite{Ste-Vie:1999:PHYSRL:,Kot-Stu-Tor:2008:CLAQG:,Stu-Kot:2009:GRG:,Sche-Stu:2009:GenRelGrav:} are assumed in models explaining the HF QPOs in the black hole systems. Alternatively, the oscillations of tori \cite{Rez-Yos-Zan:2003:MONNRS:} or tilted oscillating discs \cite{Kato:2008:PASJ:} are considered as explanation of the HF QPOs. The frequencies of the oscillating tori and tilted discs are related to the orbital and epicyclic geodesic frequencies, if the oscillations are governed mainly by the gravity of the black hole \cite{Stu-Kot-Tor:2013:ASTRA:}. 

It was suggested in \cite{Klu-Abr:2000:ASTROPH:} that resonances in oscillations of Keplerian accretion discs have to be relevant. This assumption seems to be confirmed due to the properties of the HF QPOs observed in the three Galactic microquasars GRS 1915+105, XTE 1550-564, GRO 1655-40, where the $3:2$ frequency ratio has been observed \cite{Tor-etal:2005:ASTRA:}. Note that there has been observed more complex frequency pattern in the microquasar GRS 1915+105, containing five frequencies, and more complex theory has to be involved in order to explain the whole frequency set \cite{Stu-etal:2005:PHYSR4:,Stu-Sla-Tor:2007:ASTRA:}. Here we focus our attention to the twin HF QPOs demonstrating the $3:2$ frequency ratio. 

%-------------------------------------------------------------------------%
\begin{figure*}
\includegraphics[width=\hsize]{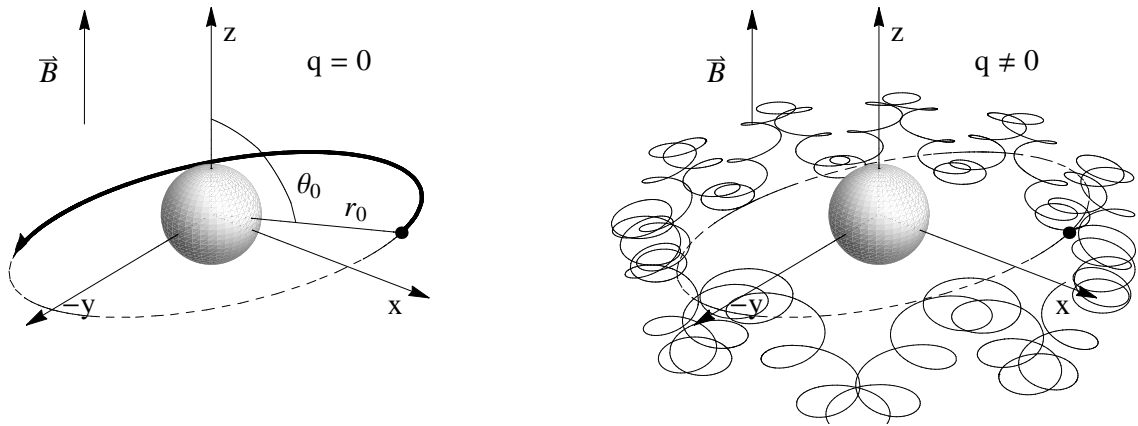}
\caption{
\label{INIorbits}
Quasi-periodic motion of a charged test particle around a Schwarzschild black hole immersed in an external uniform magnetic field (right) compared to the motion of an electrically neutral test particle moving along a circular orbit around the black hole (left). The charged particle starts its motion at the point on the circular geodesic of the uncharged particle. If the energy and angular momentum of the charged particle is slightly modified with respect to those of the circular geodesic, quasi-harmonic oscillations can occur. 
}
\end{figure*}
%-------------------------------------------------------------------------%

Unfortunately, the observed frequencies of the $3:2$ HF QPOs in the three microquasars cannot be explained by a unique model based on the frequencies of the geodesic epicyclic motion, if the limits on the mass and spin of the black holes are taken into account \cite{Tor-etal:2011:ASTRA:,Bam:2012:JCAP:}. In the present paper, we devote our attention to the possibility to explain the observed $3:2$ HF QPOs in the three microquasars due to charged particles oscillating around stable circular geodesics in combined gravitational and electromagnetic fields of magnetized black holes. Observational evidence for existence of a magnetic field around astrophysical black holes can be found in \cite{Koo-Bic-Kun:1999:PASA:,Mil-etal:2006:NATUR:}. 

The role of the magnetic field in the processes taking place in the black hole surroundings is relevant due to several reasons. Most of the observed black hole candidates have an accretion disc constituted from  conducting plasma which dynamics can generate a regular magnetic field. Recently, it has been found that the center of the Galaxy has a strong magnetic field around a supermassive black hole that is not related to an accretion disc \cite{Eat-etal:2013:NATUR:}. Therefore, black holes can be also immersed in an external, large scale electromagnetic field that can have a complicated structure in vicinity of field source, but at large distance (asymptotically) in finite element of space, its character can be simple and close to a homogeneous magnetic field. Moreover, it has been recently shown \cite{Kov-etal:2014:PHYSR4:} that a black hole located near the equatorial plane of a magnetar will be immersed in a nearly homogeneous magnetic field if distance to the magnetar is large enough. Hereafter in this paper we will concentrate our attention on this particular and simplified case of a black hole immersed in an asymptotically uniform magnetic field known as Wald solution for a magnetized black hole \cite{Wald:1974:PHYSR4:}. 

According to \cite{Pio-etal:2010:ARXIV:}, where the magnetic coupling process has been considered, in the vicinity of a stellar mass, $M\aprx10\,{M}_{\odot}$, or a supermassive, $M\aprx10^9\,{M}_{\odot}$, black hole, the estimate of the strength of the magnetic field gives, $B\aprx10^8$\,Gauss, or,  $B\aprx10^4$\,Gauss, respectively. It has to be stressed that in both cases the magnetic field does not change the gravitational background so that the metric tensor of the Schwarzschild black hole spacetime does not need a modification. In order to satisfy this test field approximation, the strength of the magnetic field in the vicinity of a black hole with the mass $M$ has to fulfil the condition \cite{Fro-Sho:2010:PHYSR4:}
\beq \label{BBB}
B << B_{\rm G}={ \frac{c^4}{G^{3/2}} {M}_{\odot}}\left(\frac{{M}_{\odot}}{M}\right)\aprx10^{19}{ \frac{{M}_{\odot}} {M} }\,{\rm Gauss}\, .
\eeq
The magnetic fields satisfying the condition of the test field approximation, $B<<B_G$, have negligible effect on the motion of neutral particles. However, for the motion of charged test particles the influence of the magnetic field can be really large. For a charged test particle with charge $q$ and mass $m$ moving in vicinity of a black hole with mass $M$ surrounded by an external asymptotically uniform magnetic field of the strength $B$, one can introduce a dimensionless quantity $b$ that can be identified as relative Lorenz force ~\cite{Fro-Sho:2010:PHYSR4:}
\beq
b=\frac{|q| B G M}{m c^4}.
\eeq
This quantity can be quite large even for weak magnetic fields due to the large value of the specific charge $q/m$~\cite{Pio-etal:2011:ASBULL:,Fro:2012:PHYSR4:}
\beq
b \aprx 4.7 \times 10^6 \left( \frac{q}{e} \right) \left( \frac{m}{m_{\rm{p}}} \right)^{-1} \left( \frac{ B } {10^8\,{\rm Gauss }}\right) \left( \frac{M}{ 10\,{M}_{\odot} } \right), \label{estimb}
\eeq
where $m_{\rm{p}}$ is the proton mass and $e$ is its charge. According to the estimation (\ref{estimb}) the influence of the magnetic field on the motion of charged particles cannot be neglected even for weak magnetic fields.

Motion of charged test particles in combined gravitational and electromagnetic fields around black holes has been studied in a large variety of papers both for the Reissner-Nordstrom and Kerr-Newman solutions of the Einstein-Maxwell equations \cite{Ruf:1973:BlaHol:,Pug-Que-Ruf:2011:PHYSR4:,Bic-Stu-Bal:1989:BAC:,Bal-Bic-Stu:1989:BAC:,Stu-Kot:2009:GRG:} and black holes with external test electromagnetic fields \cite{Prasanna:1980:RDNC:,Ali-Gal:1981:GRG:,Fro-Sho:2010:PHYSR4:,Fro:2012:PHYSR4:,Zah-etal:2013:PHYSR4:,Pre:2004:PHYSR4:,Abd-etal:2013:PHYSR4:,Bak-etal:2010:CLAQG:}.
Existence of the off-equatorial circular motion of charged particles in the combined electromagnetic and gravitational fields has been considered in \cite{Kov-Stu-Kar:2008:CLAQG:,Kov-etal:2010:CLAQG:,Kop-etal:2010:APJ:}. Possible existence of the off-equatorial levitating toroidal configurations has been also demonstrated in \cite{Cre-etal:2013:ApJS:}. 
The astrophysically relevant energetic mechanisms such as the "magnetic" Penrose process or the Blandford-Znajek mechanism for black holes surrounded by a toroidal electric current have been studied in \cite{Wagh-Dadhich:1989:PR:,Li-Xin:2000:PHYSR4:}. 

The motion of charged test particles can be treated using Hamilton equations and related effective potential. For black holes with an external magnetic field, the motion is generally of chaotic character, but in vicinity of the local minima of the effective potential the motion is simplified and corresponds to linear harmonic oscillations or quasi-harmonic oscillation with frequencies close to those of the harmonic motion. The quasi-harmonic oscillations around a stable equilibrium location can be relevant for explanation of the HF QPOs observed in the three microquasars GRS 1915+105, XTE 1550-564, and GRO 1655-40 \cite{Tor-etal:2005:ASTRA:,Stu-Kol:2014:PHYSR4:}. 

In this paper we consider motion of charged test particles around a \Schw{} black hole immersed in an external uniform magnetic field. We look especially for existence and properties of the harmonic or quasi-harmonic oscillations of charged particles in the magnetized black hole backgrounds, see Fig. \ref{INIorbits}.
Many individual charged particles will probably not create complete accretion disc, since the fluid element of the accretion disk will be largely charge neutral, but the charged particle epicyclic motion can still be used for dynamics of ionized blob structures created by instabilities or by irradiation in otherwise neutral accretion disk \cite{Stu-Kot-Tor:2013:ASTRA:}.

Throughout the paper, we use the spacetime signature $(-,+,+,+)$, and the system of geometric units in which $G = 1 = c$. However, for expressions having an astrophysical relevance we use the speed of light explicitly. Greek indices are taken to run from 0 to 3.

%%%%%%%%%%%%%%%%%%%%%%%%%%%%%%%%%%%%%%%%%%%%%%%%%%%%%%%%%%%%%%%%%%%%%%%%%%%
\section{Charged particle dynamics}
%%%%%%%%%%%%%%%%%%%%%%%%%%%%%%%%%%%%%%%%%%%%%%%%%%%%%%%%%%%%%%%%%%%%%%%%%%%
Using the general Hamiltonian formalism, we describe dynamics of a charged particle with charge $q\neq~0$ in the vicinity of the Schwarzschild black hole embedded in external asymptotically uniform magnetic field. The equations of motion of neutral particles can be obtained by taking $q=0$. 

The line element of the \Schw{} black hole spacetime with mass $M$ reads 
\beq
    \d s^2 = -f(r) \d t^2 + f^{-1}(r) \d r^2 + r^2(\d \theta^2 + \sin^2\theta \d \phi^2), \label{SCHmetric}
\eeq
where the function $f(r)$ takes the form 
\beq 
		f(r) = 1 - \frac{2 M}{r}. 
\eeq
Hereafter, we put $M=1$, i.e., we use dimensionless radial coordinate $r$ (and time coordinate $t$).

The stationarity and axial symmetry of the Schwarzschild spacetime imply existence of the timelike and spacelike Killing vectors that satisfy the equation 
\beq 
	\xi_{\alpha;\beta} + \xi_{\beta;\alpha} = 0. 
\eeq
This implies that the solution for the four-vector potential $A^\mu$ representing the test electromagnetic field takes the following form \cite{Wald:1984:book:}
\beq 
	A^{\mu} = C_1 \xi^{\mu}_{(t)} + C_2 \xi^{\mu}_{(\phi)}.
\eeq
We consider the case of magnetic field which is uniform at the spatial infinity, having strength $B$ there. The field is oriented perpendicularly to the equatorial plane of the black hole spacetime. Then the four-vector potential takes the form 
\beq
A^{\alpha} = \frac{B}{2} \, \xi^{\alpha}_{(\phi)}. \label{EMpotenc2}
\eeq
The commuting Killing vector $\xi_{(\phi)} = \partial / \partial \phi$ generates rotations around the symmetry axis. Consequently, the only nonzero covariant component of the potential of the electromagnetic field takes the form \cite{Wald:1984:book:}
\beq
A_{\phi} = \frac{B}{2} \, g_{\phi\phi}  = \frac{B}{2} \, r^2 \sin^2 \theta. \label{aasbx}
\eeq
The charged test particle motion is described by the Lorentz equation
\beq 
	m \frac{\d u^\mu}{\d \tau} = q F^{\mu}_{\nu} u^{\mu}, 
\eeq
where $u^{\mu}$ is the four-velocity of the particle with the mass $m$ and charge $q$, normalized by the condition $u^{\mu} u_{\mu} = -1$, $\tau$ is the proper time of the particle, and $F_{\mu \nu} = A_{\nu,\mu} - A_{\mu,\nu}$ is the antisymmetric tensor of the electromagnetic field.

%%%%%%%%%%%%%%%%%%%%%%%%%%%%%%%%%%%%%%%%%%%%%%%%%%%%%%%%%%%%%%%%%%%%%%%%%%%
\subsection{Hamiltonian and effective potential}
%%%%%%%%%%%%%%%%%%%%%%%%%%%%%%%%%%%%%%%%%%%%%%%%%%%%%%%%%%%%%%%%%%%%%%%%%%%

The Hamiltonian for the charged particle motion can be written in the form \cite{Wald:1984:book:}
\beq
  H_{\rm p} =  \frac{1}{2} g^{\alpha\beta} (\cp_\alpha - q A_\alpha)(\cp_\beta - q A_\beta) + \frac{1}{2} \, m^2
  \label{particleHAM},
\eeq
where the kinematical four-momentum $\p^\mu = m u^\mu$ is related to the generalized (canonical) four-momentum $\cp^\mu$ by the relation
\beq
 \cp^\mu = \p^\mu + q A^\mu, \label{particleMOM}
\eeq
that satisfy the Hamilton equations in the form 
\beq
 \frac{\d \x^\mu}{\d \af} \equiv \p^\mu = \frac{\partial H}{\partial \cp_\mu}, \quad
 \frac{\d \cp_\mu}{\d \af} = - \frac{\partial H}{\partial \x^\mu}. \label{Ham_eq}
\eeq
The affine parameter $\af$ of the particle is related to its proper time $\tau$ by the relation $\af=\tau/m$.

%-------------------------------------------------------------------------%
\begin{figure*}
\includegraphics[width=\hsize]{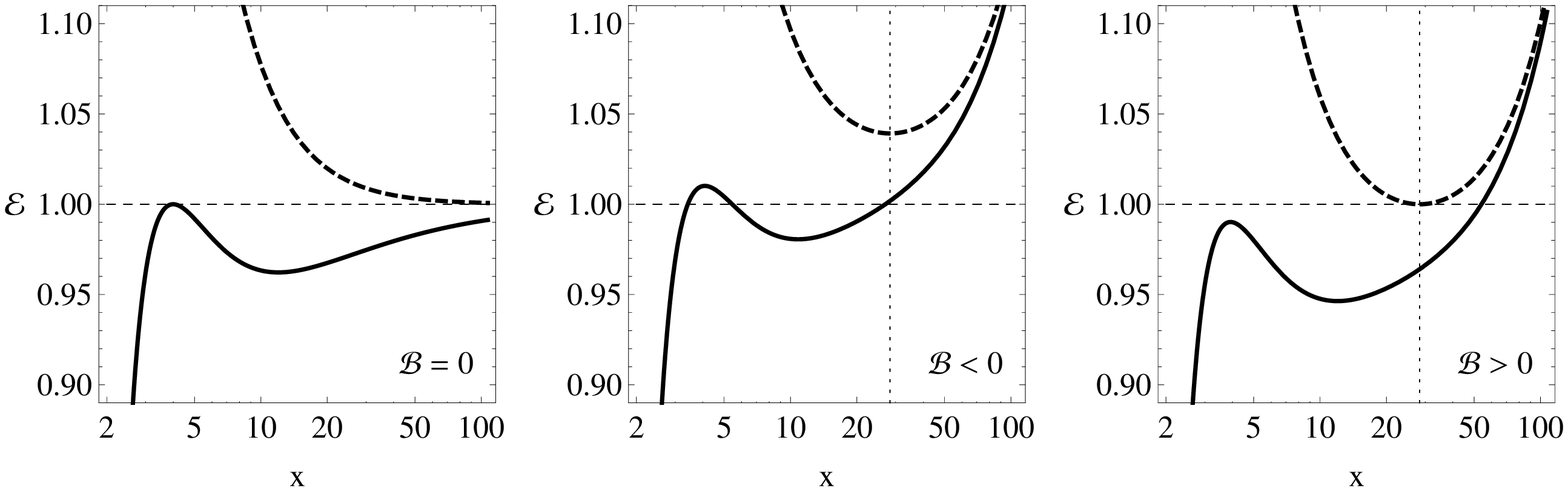}
\caption{
\label{figVeff}
Sections ($y=0, z=$const.) of the effective potential $V_{\rm eff} (x,y,z;\cl,\cb)$ taken at the equatorial plane $z=0$ (solid curves) and at the spatial infinity $z\rightarrow\infty$ (dashed curves) for a charged test particle moving around a \Schw{} black hole immersed in an uniform magnetic field. We compare uncharged case, $\cb=0$ (left), with negative, $\cb<0$ (middle), and positive, $\cb>0$ (right), configurations. Values $\cl=4$ and $\cb=\pm~0.005$ are used for the graphs; the minima of $V_{\rm eff} (x,z\rightarrow\infty)$ (dotted vertical line) is at $r=20\sqrt{2}$. 
The influence of the external magnetic field becomes decisive for large values of coordinate $x$, for small values of $x$, the role of the magnetic field is suppressed by gravity in the equatorial plane and its vicinity, and by the centrifugal forces related to the angular momentum for $z\rightarrow\infty$, see eq. (\ref{VeffCharged})
}
\end{figure*}
%-------------------------------------------------------------------------%

Due to the symmetries of the Schwarzschild spacetime (\ref{SCHmetric}) and the asymptotically uniform magnetic field (\ref{aasbx}), one can easily find the conserved quantities that are the energy and the axial angular momentum of the particle and can be expressed as 
\bea
 E &=& - \cp_t = m f(r) \frac{\d t}{\d \tau}, \\
 L &=& \cp_\phi = m r^2 \sin^2\theta \left(\frac{\d \phi}{\d \tau} + \frac{q B}{2m} \right). \label{angmom}
\eea

The dynamical equations for the charged particle motion in the Cartesian coordinates can be found by the coordinate transformations 
\beq
 x = r \cos(\phi) \sin(\theta),\, y = r \sin(\phi) \sin(\theta),\, z = r \cos(\theta). \label{Coord}
\eeq

Introducing for convenience the specific parameters, energy $\ce$, axial angular momentum $\cl$, and magnetic parameter $\cb$, by the relations \cite{Fro-Sho:2010:PHYSR4:}
\beq
\ce = \frac{E}{m}, \quad \cl = \frac{L}{m}, \quad \cb = \frac{q B}{2m},
\eeq
one can rewrite the Hamiltonian (\ref{particleHAM}) in the form
\beq 
H = \frac{1}{2} f(r) \p_r^2 + \frac{1}{2r^2} \p_\theta^2  + \frac{1}{2} \frac{m^2}{f(r)} \left[ V_{\rm eff}(r,\theta) - \ce^2 \right], \label{HamHam} 
\eeq
where $V_{\rm eff}(r,\theta; \cl,\cb)$ denotes the effective potential given by the relation 
\beq 
V_{\rm eff} (r,\theta) \equiv f(r) \left[1+\left(\frac{\cl}{r \sin{\theta} } - \cb\, r \sin{\theta}\right)^2\right]. \label{VeffCharged} 
\eeq
The terms in the parentheses corresponds to the central force potential given by the specific angular momentum $\cl$, and electromagnetic potential energy given by the magnetic parameter $\cb$.

The effective potential (\ref{VeffCharged}) shows clear symmetry $(\cl,\cb)\leftrightarrow(-\cl,-\cb)$ that allows to distinguish the following two situations 
\begin{itemize}
\item[-] {\it minus configuration}, here $\cl>0, \cb<0$  (equivalent to $\cl<0, \cb>0$) - magnetic field and angular momentum parameters have opposite signs and the Lorentz force is attracting the charged particle to the $z$-axis, towards the black hole.
\item[+] {\it plus configuration}, here $\cl>0, \cb>0$ (equivalent to $\cl<0, \cb<0$) - magnetic field and angular momentum parameters have the same signs and the Lorentz force is repulsive, acting outward the black hole.
\end{itemize}
The positive angular momentum of a particle $\cl>0$ means that the particle is revolved in the counter-clockwise motion around the black hole, see Fig. \ref{INIorbits}. If charge of the particle is taken to be positive $q>0$, the minus configuration $\cb<0$ corresponds to the vector of the magnetic field $\vec{B}$ pointing downwards, while plus configuration $\cb>0$ corresponds to the vector of the magnetic field $\vec{B}$ pointing upwards the $z$-axis.

The charged particle motion is limited by the energetic boundaries given by
\beq
 \ce^2 = V_{\rm eff} (r,\theta; \cl,\cb). \label{MotLim}
\eeq
Let us properly investigate the features of the effective potential (\ref{VeffCharged}) represented in Fig. \ref{figVeff} that enables us to demonstrate general properties of the charged particle dynamics, avoiding the necessity to solve the equations of motion \cite{Fro-Sho:2010:PHYSR4:}. 

The axial symmetry of the background of the combined gravitational and magnetic fields implies independence of the effective potential $V_{\rm eff}$ on the coordinate $\phi$ which allows us to examine $V_{\rm eff}(r,\theta)$ as a 2D function of variables $r,\theta$ or Cartesian $x,z$ coordiantes (\ref{Coord}). The effective potential is positive outside the black hole horizon, and diverges at the horizon $r=2$. The region within the horizon and divergent points is excluded from our investigation.

The stationary points of the effective potential $V_{\rm eff}(r,\theta)$ function, where maxima or minima can exist, are given by the equations 
\beq
  \partial_r V_{\rm eff}(r,\theta;\cl,\cb) = 0, \quad \partial_\theta V_{\rm eff}(r,\theta;\cl,\cb) = 0.  \label{extrem}
\eeq
The second of the extrema equations (\ref{extrem}) gives $\theta=\pi/2$. In another words, all extrema of the $V_{\rm eff}(r,\theta);\cl,\cb$ function are located in the equatorial plane and there is no off-equatorial extremum for the charged particles orbiting the \Schw{} BH immersed in the uniform magnetic field, in contrast to the dipole magnetic field case \cite{Kov-Stu-Kar:2008:CLAQG:}. 

The first extrema equation of (\ref{extrem}) leads to a polynomial equation of the fifth order in the radial coordinate 
\beq
  \cl^2 (r-3) + 2 \cl \cb r^2 - \cb^2 r^4 (r-1) - r^2 = 0, \label{dveffdr}
\eeq
which has generally five complex roots. Real roots of (\ref{dveffdr}) above the black hole horizon determine maxima, minima and inflex points of the $V_{\rm eff}(r,\theta=\pi/2;\cl,\cb)$ function. Such extrema give stable (minima) and unstable (maxima) equilibrium positions for the circular particle motion, i.e. stable or unstable circular orbits. The inflex points give the marginally stable circular orbits. 

\begin{figure*}
\includegraphics[width=\hsize]{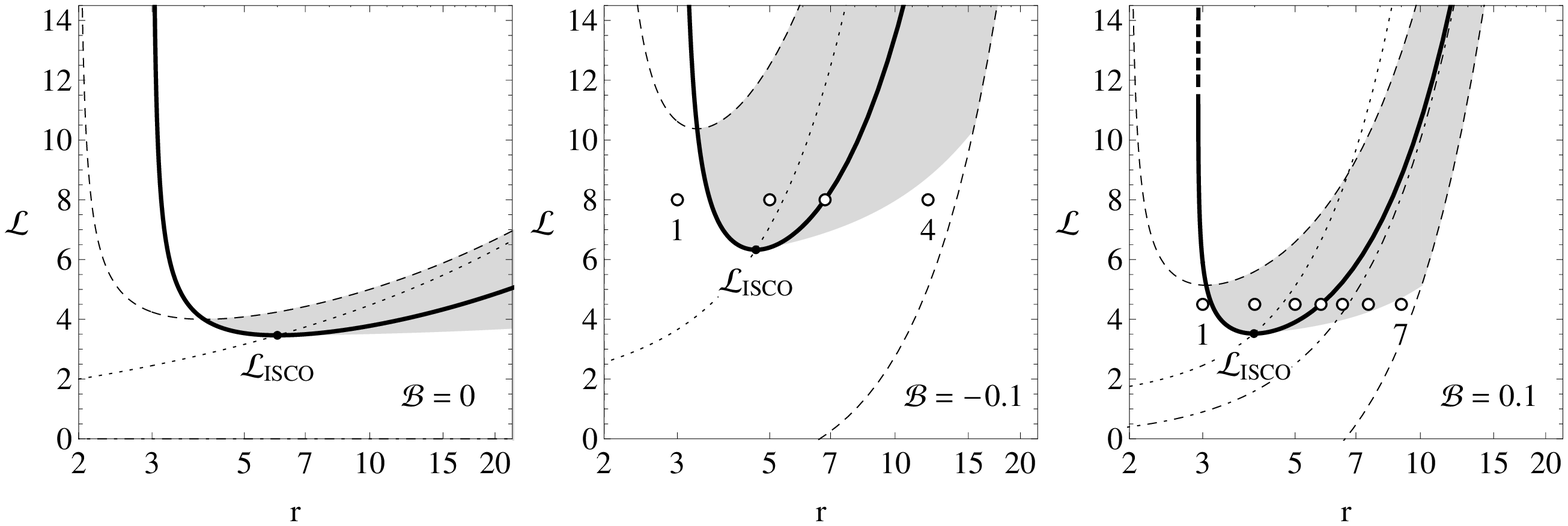}
\caption{\label{rLfig}
Circular orbits of charged particles. The extrema function $\cl_{\rm E\pm}(r)$ (thick) determines local extrema of the effective potential $V_{\rm eff}(r,\theta)$, i.e., the circular orbits in the equatorial plane. $\cl_{\rm ISCO}$ governs the innermost stable circular orbit (relative to radial perturbations), the stable (unstable) circular orbits are located at $r>r_{\rm ISCO}$ ($r<r_{\rm ISCO}$). The functions $\cl_{\rm L1}(r), \cl_{\rm L2}(r)$ (dashed) govern the trapped states. The function $\cl_{\rm E(ex)}(r;\cb)$ (dotted) governs the extrema of the $\cl_{\rm E\pm}(r)$ function. The negative branch of the extrema function, $\cl_{\rm E-}(r)$ (thick dashed curve), and the curve $\cl_{*}(r;\cb)$ (dot-dashed) governing the existence of the trajectories with curls, can both exist only for positive values of the magnetic parameter, $\cb>0$.  
Regions where trapped states (bounded orbits only) can exist are shaded.
All curves are given for neutral $\cb=0$ (left), negative $\cb<0$ (middle) and positive $\cb>0$ (right) values of the charged test particle magnetic parameter.
Numbers denote the examples of different types of trajectories illustrated in Figs. \ref{traM} and \ref{traP}.
}
%\end{figure*}
%-------------------------------------------------------------------------%
%\begin{figure*}
\includegraphics[width=\hsize]{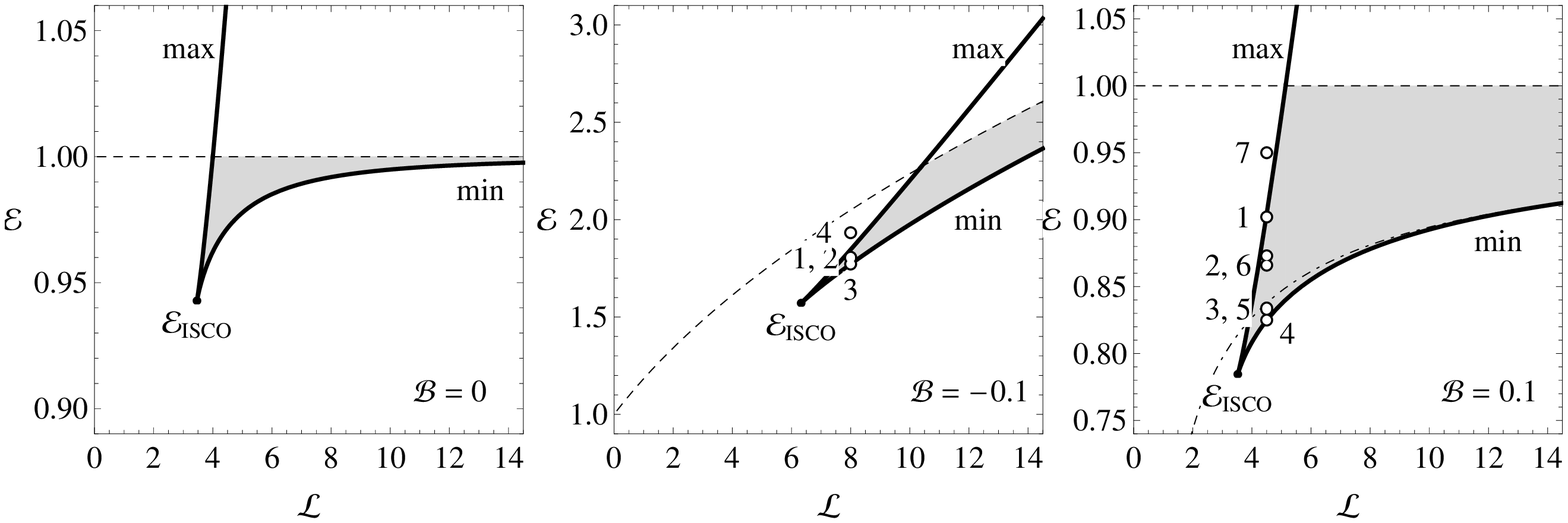}
\caption{
\label{LEfig}
Energetics of the circular orbits. Local extrema of the effective potential $V_{\rm eff}(r,\theta)$ determine the circular orbits in the equatorial plane and are given as functions of the specific angular momentum $\cl$. Thick solid curves correspond to the maxima (unstable orbits) and minima (stable orbits) of the effective potential $V_{\rm eff}(r,\theta)$. Minimum of the effective potential for the flat spacetime $\ce_{\rm flat(min)}$ (governing escape to infinity) is represented by the dashed line. For $\cb>0$, the curly motion is allowed for particles with parameters above the function $\ce_{*}(\cl)$ (dot-dashed line). Regions where the trapped states can exist are shaded. 
Numbers denote the examples of different types of trajectories illustrated in Figs. \ref{traM} and \ref{traP}.
}
\end{figure*}

The equation (\ref{dveffdr}) is quadratic with respect to the specific angular momentum $\cl$ and hence the circular orbits can be determined by the relation 
\beq
 \cl = \cl_{\rm E\pm}(r;\cb) \equiv \frac{-\cb r^2 \pm r F}{r-3}, \label{Lcirc}
\eeq
where we have introduced new function
\beq
 F(r;\cb) = \sqrt{ \cb^2 r^2 (r - 2)^2 + r - 3 }. \label{Ffunction}
\eeq
The positive branch of (\ref{Lcirc}), the solution $\cl_{\rm E+}>0$, can exist at the whole $r>2$ region and determines both the stable and unstable circular orbits; for the plus configurations with $\cb>0$, both functions $\cl_{\rm E\pm}>0$ can exist in the $2<r<3$ region, see Fig. \ref{rLfig}. The negative branch of (\ref{Lcirc}), the solution $\cl_{\rm E-}$, is responsible for maxima of the effective potential $V_{\rm eff}$ only.

%-------------------------------------------------------------------------%
%-------------------------------------------------------------------------%

%-------------------------------------------------------------------------%
\begin{figure*}
\includegraphics[width=\hsize]{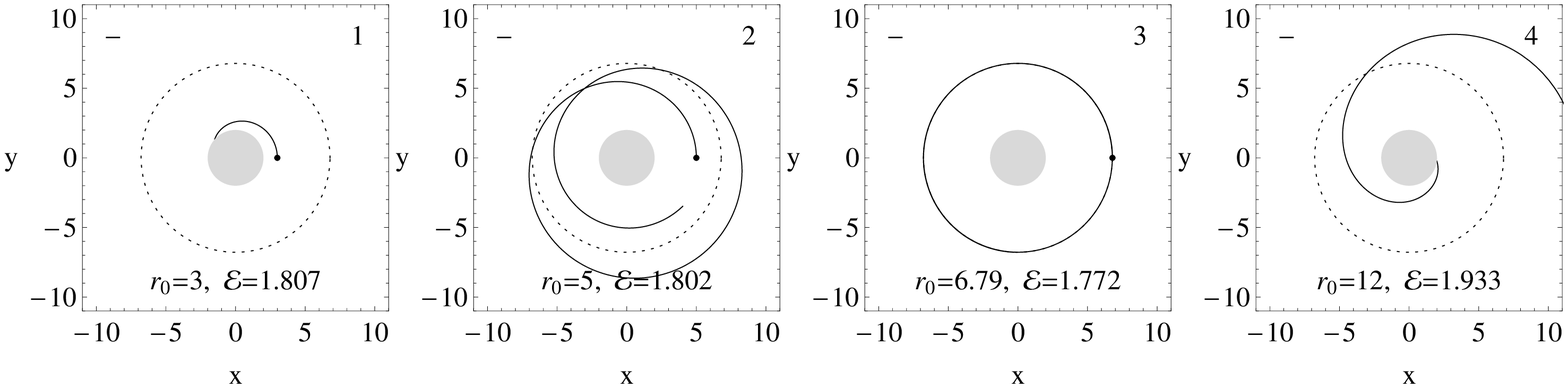}
\caption{\label{traM}
List of charged test particle equatorial trajectories with specific angular momenta $\cl=8$ around a \Schw{} black hole immersed in an uniform magnetic field -- the magnetic parameter $\cb=-0.1$, see Fig. \ref{rLfig}. (middle). The particle starts from different initial radii $r$ clarifying different shapes of its trajectory. Dotted circle represents stable circular orbit for given specific angular momenta $\cl$.
}
\end{figure*}
%-------------------------------------------------------------------------%
\begin{figure*}
\includegraphics[width=\hsize]{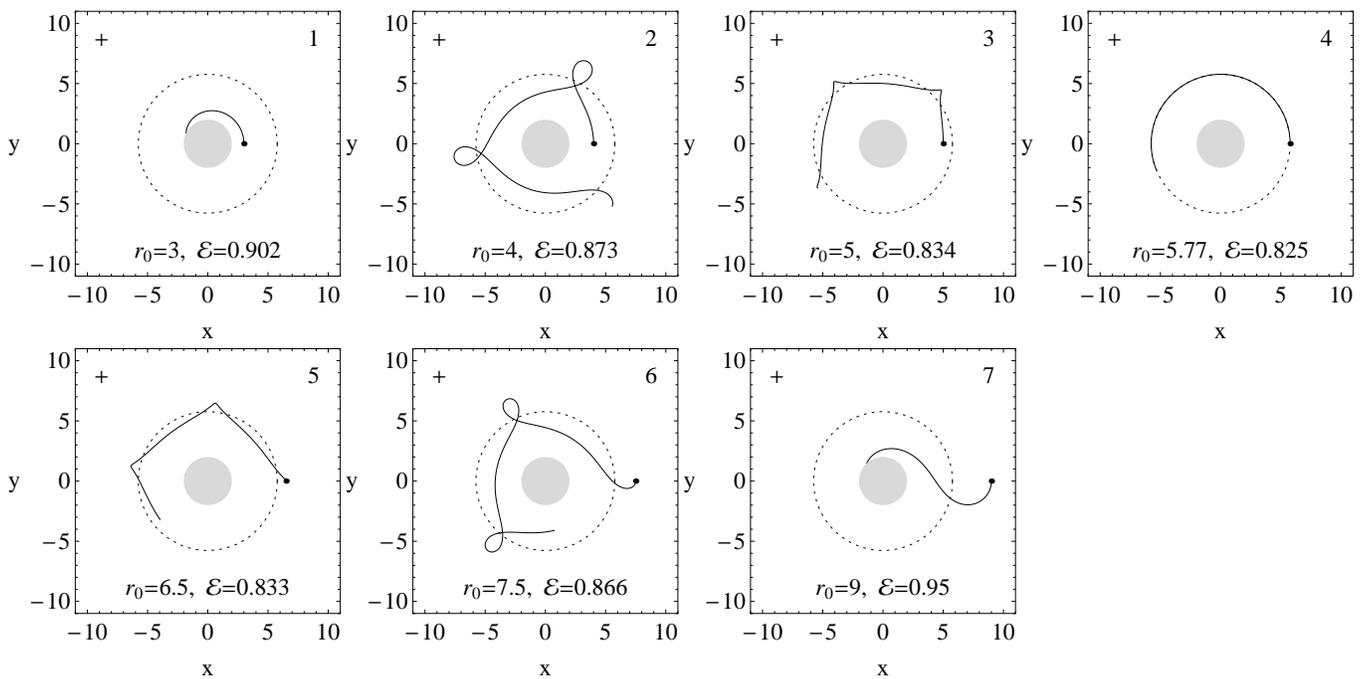}
\caption{\label{traP}
Curling trajectories. List of charged test particle equatorial trajectories with specific angular momenta $\cl=4.5$ around a \Schw{} black hole immersed in an uniform magnetic field -- the parameter $\cb=0.1$, see Fig. \ref{rLfig}. (right). 
}
\end{figure*}
%-------------------------------------------------------------------------%

The local extrema of the $\cl_{\rm E\pm}(r)$ function determine the innermost stable circular orbits (ISCO), their radius, angular momentum, and energy. They are determined by the condition $\partial^{2}_{r} V_{\rm eff}(r,\theta;\cl,\cb) = 0$ that implies the relation 
\beq
 \cl_{\rm E(ex)}(r;\cb) =  - 2 \cb r + G,
\eeq
where 
\beq
 G(r;\cb) = \sqrt{\cb^2 r^2 \left(5 r^2-4 r+4\right) + 2r}. \label{Gfunction}
\eeq
The ISCO is located at the point of intersection of the functions $\cl_{\rm E(ex)}$ and $\cl_{\rm E+}$, see Fig. \ref{rLfig}., which leads to the condition for the ISCO radius given in the form 
\beq
6-r + 2 \cb (r-6) G - 2 \cb^2 r \left(2 r^3 - 9 r^2 + 8 r - 12\right) = 0. \label{ISCO}
\eeq

Contrary to the uncharged test particle motion with $\cb=0$, the motion of charged particles in the uniform magnetic field is always bounded in the $r$ direction near the equatorial region $\theta\aprx\pi/2$, because of the term ~$\cb^2 r^2$ in effective potential (\ref{VeffCharged}), which unlimitedly grows with $r~\rightarrow~\infty$. However, the energetic boundary (\ref{MotLim}) for the charged particle motion can be open in the polar direction of the black hole spacetime, the $z$ direction, making the charged particles able to escape to infinity along the $z$ axis. The energetic condition for the particles which can escapes to infinity reads
\beq
 \ce \geq \ce_{\rm flat(min)} = 
\Big\{ 
\begin{array}{l @{\quad} c @{\quad} l} 
1 & \textrm{for} & \cb \geq 0 \\ 
\sqrt{1 + 4 \cb \cl} & \textrm{for} & \cb < 0 \\ 
\end{array} \label{escape}
\eeq
where $\ce_{\rm flat(min)}$ is the minimal energy of the charged particle at infinity, see Fig. \ref{LEfig}.

If the energetic boundary (\ref{MotLim}) forms a closed curve, the motion of charged particles is trapped, and their trajectory is restricted to a toroidal-like region governed by the effective potential. The condition for formation of the closed ``lake'' like energy boundaries that are related to the local minima of the effective potential $V_{\rm eff}(x,z)$ obviously reads 
\beq
	\ce < \ce_{\rm flat(min)} \label{eq_Econ}.
\eeq
One can deduce that the trapped states of the oscillating charged particle can exist, if
\beq
 \cl_{\rm L1} < \cl < \cl_{\rm L2} \label{eqTrap}
\eeq
where the so called ``lake'' angular momentum functions $\cl_{\rm L1}(r; \cb)$ and $\cl_{\rm L2}(r; \cb)$, are the solutions of the equality condition $\ce = \ce_{\rm flat(min)}$. For $\cb \geq 0$, the functions $\cl_{\rm L1}(r; \cb)$ and $\cl_{\rm L2}(r; \cb)$ take the form 
\beq
 \cl_{L1,L2} = \frac{\cb r^2 (r-2) \pm r \sqrt{2(r-2)}}{r-2}, \label{fceL1L2p} 
\eeq
while for $\cb < 0$, they take the form 
\beq
 \cl_{L1,L2} = \frac{-\cb r^2 (r+2) \pm r \sqrt{2(r-2 + 4 \cb^2 r^3 )}}{r-2}. \label{fceL1L2m}
\eeq
The functions $\cl_{\rm L1}(r), \cl_{\rm L2}(r)$ are represented in Fig. \ref{rLfig}, for $\cb=0$ and $\cb=\pm~0.1$. 

The energy of the circular orbits of the charged particles, given by the local minima and maxima of the effective potential, is represented in Fig. \ref{LEfig}. Typical sequences of the related charged particle trajectories, for given specific angular momentum $\cl$, are represented in Figs. \ref{traM} and \ref{traP}. 

\subsection{Curled trajectories}

Using the conservation of the angular momentum (\ref{angmom}) of the motion of a charged particle around \Schw{} black hole in uniform magnetic field, the equation of motion (\ref{Ham_eq}) for the axial coordinate $\phi$ can be written in the form
\beq
 \dot{\phi} = \frac{\cl}{r^2} - \cb.
\eeq
For uncharged particles, the case $\cb=0$ ($\cl>0,r>0$), the right side of the above equation is always positive and hence the coordinate $\phi$ can only increase. But for positive values of the electromagnetic interaction parameter, $\cb > 0$, the coordinate $\phi$ can decrease, if
\beq
 \cl > \cl_{*}(r;\cb) \equiv \cb r^2.
\eeq
This condition can be expressed in terms of the energy condition \cite{Fro-Sho:2010:PHYSR4:}
\beq
 \ce > \ce_{*}(\cl;\cb) \equiv \sqrt{1- 2\sqrt{\cb/\cl}}.
\eeq
Decrease of the coordinate $\phi$ during the charged particle motion leads to the epicyclic motion with curled trajectories. The "curling" functions, $\cl_{*}(r)$ and $\ce_{*}(\cl)$, are illustrated in Figs. \ref{rLfig} and \ref{LEfig}, respectively. The transition from a trajectory without curls to those demonstrating the curly character is illustrated for a constant specific angular momentum $\cl$ in Fig. \ref{traP}. Note that the curled trajectories are the most illustrative example of the influence of the magnetic field. 

%-------------------------------------------------------------------------%
\begin{figure*}[t]
\subfigure[\label{OrbitShpA} \quad collapse]{\includegraphics[width=0.24\hsize]{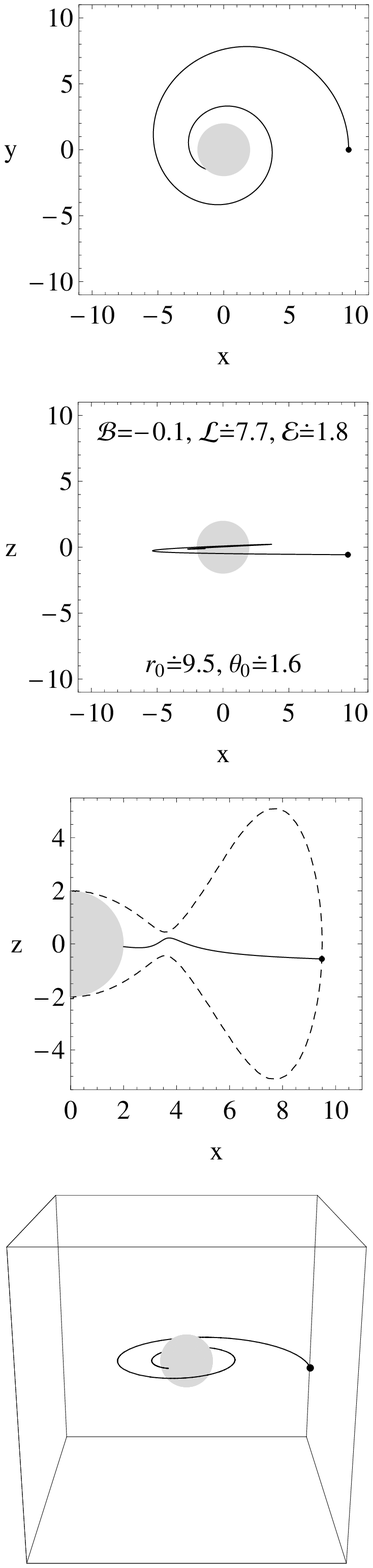}}
\subfigure[\label{OrbitShpB} \quad bounded]{\includegraphics[width=0.24\hsize]{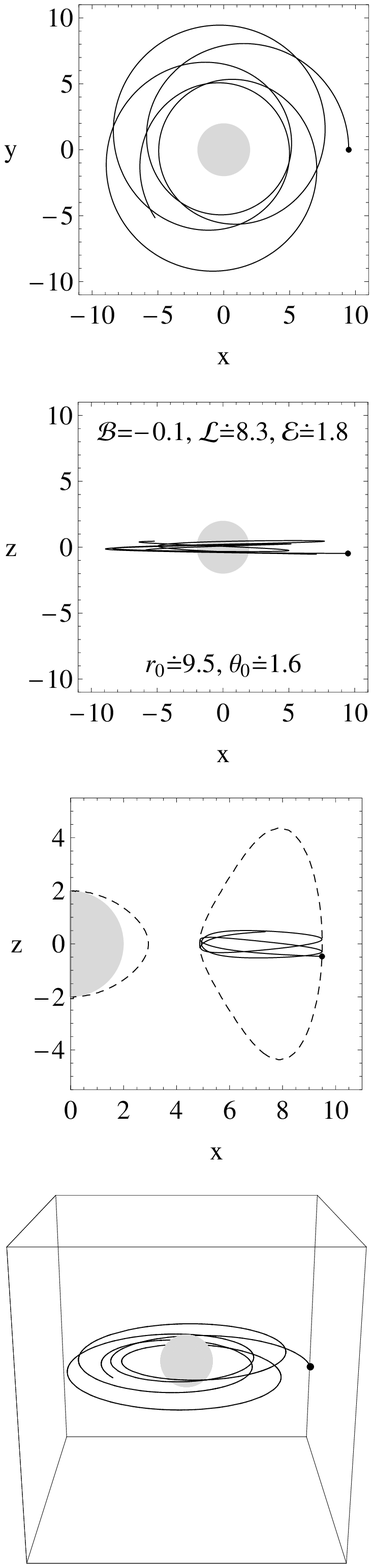}}
\subfigure[\label{OrbitShpC} bounded with curls]{\includegraphics[width=0.24\hsize]{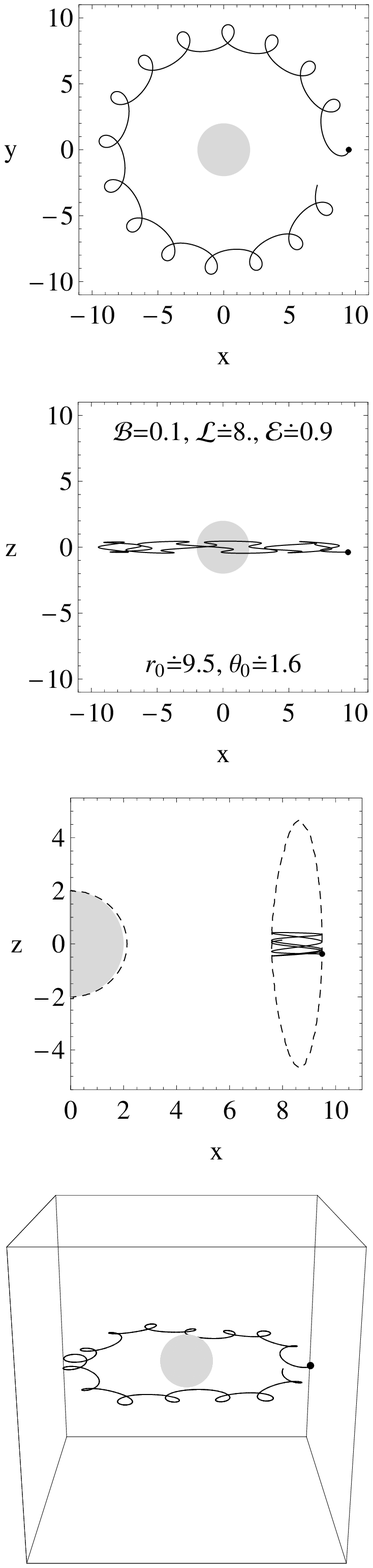}}
\subfigure[\label{OrbitShpD} \quad escape]{\includegraphics[width=0.24\hsize]{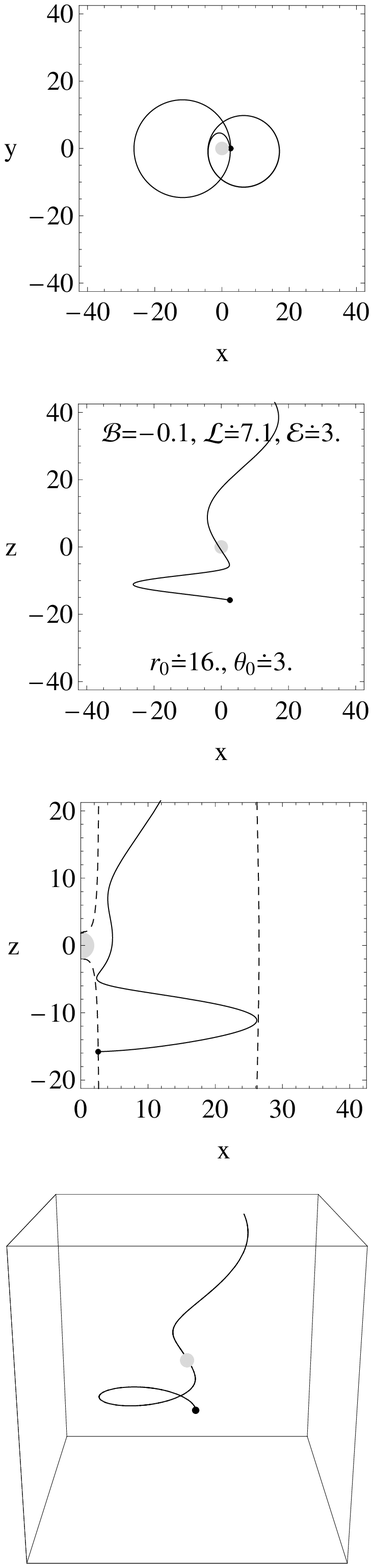}}
\caption{\label{OrbitShp}
Typical trajectories of charged test particles (solid curves) orbiting a \Schw{} black hole (shaded circle) immersed in an uniform magnetic field. The value of the electromagnetic interaction parameter is chosen as $\cb=\pm0.1$. We give $x$-$y$ and $x$-$z$ sections (first and second line) of complete 3D $x$-$y$-$z$ particle trajectory (fourth line). Due to the conservation of the particle specific energy $\ce$ and specific angular momentum $\cl$, the trajectory in 4D configuration space $(t,x,y,z)$ can be represented in 2D $x$-$z$ graph (third line), where we also plotted the boundary of the particle motion given by the effective potential (dashed curve).}
\end{figure*}
%-------------------------------------------------------------------------%

%%%%%%%%%%%%%%%%%%%%%%%%%%%%%%%%%%%%%%%%%%%%%%%%%%%%%%%%%%%%%%%%%%%%%%%%%%%
\subsection{Trajectories of charged particles}
%%%%%%%%%%%%%%%%%%%%%%%%%%%%%%%%%%%%%%%%%%%%%%%%%%%%%%%%%%%%%%%%%%%%%%%%%%%

Generally, the motion of a charged particle around a \Schw{} black hole immersed in an external uniform magnetic field (Wald solution) is chaotic \cite{Kop-etal:2010:APJ:,Kop-Kar:2014:APJ:}. However, trajectories of charged particles, which are close to the minima of the effective potential corresponding to the stable circular orbits, still have a regular (harmonic) character \cite{Kop-etal:2010:APJ:}. Moreover, the trajectories entirely located in the equatorial plane are also regular, and the chaotic behaviour occurs due to varying the inclination angle $\theta_0$ from the equatorial plane \cite{Kop-Kar:2014:APJ:}.  Here we will focus on the regular (harmonic), or quasi-harmonic bounded orbits located close to the equatorial plane, with initial inclination angle $\theta_0\aprx\pi/2$, see Fig. \ref{INIorbits}.

Due to the behaviour of the effective potential $V_{\rm eff}(x,z)$, one can distinguish four different types of the energetic boundary (\ref{MotLim}). The differing energetic boundaries can be seen as dashed curves in the third line in Fig. \ref{OrbitShp}. The {\it first case} corresponds to non existing inner and outer boundary - the particle can be captured by the black hole or escape to infinity, see Fig. \ref{OrbitShp}(d). The {\it second case} corresponds to the situation with an outer boundary - the charged particle must be captured by the black hole, see Fig. \ref{OrbitShp}(a). The {\it third case} corresponds to the situation when both inner and outer boundaries exist - the charged particle is trapped in some region forming a toroidal region around the black hole horizon, corresponding to a potential ``lake'' around the black hole, see Figs. \ref{OrbitShp}(b,c). The {\it fourth case} corresponds to the existence of the inner boundary - the particle cannot fall into the black hole, but it must escape to infinity. 

We can also distinguish four different types of the charged particle regular motion. We start with the collapse trajectories which end at the horizon (we discuss motion only above horizon, at $r>2$). Undoubtedly, the collapsed trajectories can appear only in the first and second cases of the energetic boundary that are to enable particles to reach the horizon. An example of such collapsed trajectory is represented in Fig. \ref{OrbitShp}(a). 
Then there are escaped trajectories, which end at the spatial infinity, $z\rightarrow\infty$, evolving along the magnetic field lines. Escaped trajectories can appear only in the first and fourth cases of the energetic boundary, enabling the possibility to escape the strong gravity black hole region along the $z$-axis. An example of such trajectory can be found in Fig. \ref{OrbitShp}(d). The escape along the $z$-axis appears to be uniquely related to the charged particle motion in the uniform magnetic field. 

Recall that motion of electrically neutral test particle around a \Schw{} black hole is always restricted to a central plane, and freely moving particle can escape to infinity only in this plane $x\rightarrow\infty$.
The last type of trajectories are the bounded trajectories -- their radial coordinate is bounded by apoapsis $r_{\rm a}$ and periapsis $r_{\rm p}$, such as $r_{\rm p}\leq~r~\leq~r_{\rm a}$. There can be two types of the bounded trajectories, namely those with and without curls -- examples of such trajectories can be found in Figs. \ref{OrbitShp}(c,b). The trajectories with curls are specific to the charged particle motion in the asymptotically uniform magnetic fields, and can not occur in the motion of uncharged particles. A special type of the bounded orbits are the stable circular orbits. We demonstrate that the ISCO of the charged particle circular orbits is always located inside the ISCO of neutral particles, located at $r=6$. 

%%%%%%%%%%%%%%%%%%%%%%%%%%%%%%%%%%%%%%%%%%%%%%%%%%%%%%%%%%%%%%%%%%%%%%%%%%%
\section{Harmonic oscillations}
%%%%%%%%%%%%%%%%%%%%%%%%%%%%%%%%%%%%%%%%%%%%%%%%%%%%%%%%%%%%%%%%%%%%%%%%%%%

If a charged test particle is slightly displaced from the equilibrium position located in a minimum of the effective potential $V_{\rm eff}(r,\theta)$ at $r_0$ and $\theta_0=\pi/2$, corresponding to a stable circular orbit, the particle will start to oscillate around the minimum realizing thus epicyclic motion governed by linear harmonic oscillations. For harmonic oscillations around the minima of the effective potential $V_{\rm eff}$, the evolution of the displacement coordinates $\rr = \rr_0 + \dr,\tt = \tt_0 + \dt$ is governed by the equations
\beq
 \ddot{\dr} + \omega^2_{\mir} \, \dr = 0, \quad \ddot{\dt} + \omega^2_{\mit} \, \dt = 0,
\eeq
where dot $\dot{\aaa} = \d \aaa/\d \tau$ denotes derivative with respect to the proper time $\tau$ of the particle, and locally measured angular frequencies of the harmonic oscillatory motion are given by \cite{Wald:1984:book:}
\beq
 \omega^2_{\mir}  =  \frac{\partial^2 V_{\rm eff}}{\partial r^2}, \quad
 \omega^2_{\mit}  =  \frac{1}{r^2 f(r)} \, \frac{\partial^2 V_{\rm eff}}{\partial \theta^2}.
\eeq
For an alternative definition of the epicyclic harmonic motion see \cite{Ali-Gal:1981:GRG:}.

%-------------------------------------------------------------------------%
\begin{figure}
\includegraphics[width=\hsize]{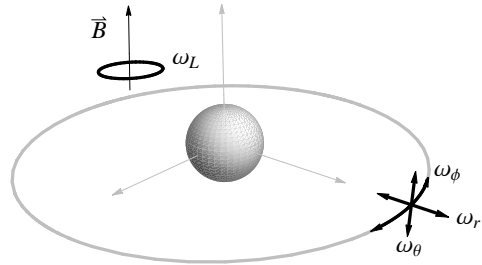}
\caption{ \label{introQPOs}
Locally measured radial (horizontal) $\omega_\mir$, latitudinal (vertical) $\omega_\mit$, Keplerian $\omega_\phi$ and Larmor $\omega_{\mil}$ angular frequencies for charged particle moving in vicinity of a stable circular orbit in the gravitational Schwarzschild field combined with the uniform magnetic field. Note that the Larmor frequency is related exclusively to the magnetic field and is thus relevant at large distance from the black hole. 
}
\end{figure}
%-------------------------------------------------------------------------%

Locally measured latitudinal (vertical) $\omega_{\mit}$ and radial (horizontal) $\omega_{\mir}$ angular frequency of the harmonic oscillations of charged particles in the combined gravitational and electromagnetic background are then given by 
\bea
\omega^2_{\mit} &=& \frac{\cl^2}{r^4}-\cb^2, \\
\omega^2_{\mir} &=& \frac{1}{(r-2) r^5} \big[ (r-2)^2 \left(\cb^2 r^4+3 \cl^2\right) \nonumber \\ 
&& \qquad\qquad\quad -2r \left( \cb r^2 -\cl \right)^2 -2r^3  \big]
\eea
where $\cl=\cl_{\rm E+}$ is the specific angular momentum at the circular orbit (\ref{Lcirc}).

Recall that there exist the third fundamental angular frequency of the epicyclic particle motion, namely the Keplerian (axial) frequency $\omega_\mip$, given by
\beq
 \omega_\mip = \frac{\d \phi}{\d \tau} = U^\phi = \frac{\cl}{g_{\phi\phi}} - \cb.
\eeq
With the uniform magnetic field itself the so called Larmor angular frequency $\omega_{\mil}$ is associated, being given by the relation 
\beq
\omega_{\mil} = \frac{q B}{m} = 2 \cb.
\eeq
Obviously, the Larmor angular frequency $\omega_{\mil}$ is not dependent on the radial coordinate $r$ and it is fully relevant in large distances from the black hole where the uniform magnetic field becomes to be crucial. For illustration of definition of the frequencies see Fig. \ref{introQPOs}. 

%-------------------------------------------------------------------------%
\begin{figure*}
\includegraphics[width=\hsize]{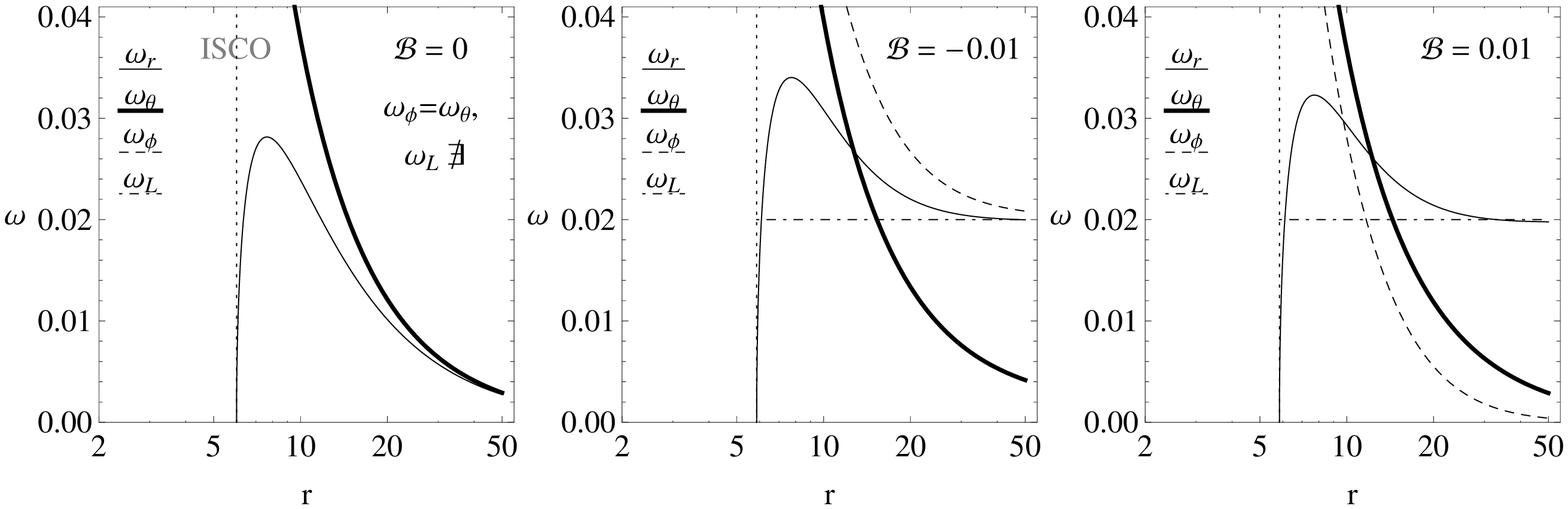}
\includegraphics[width=\hsize]{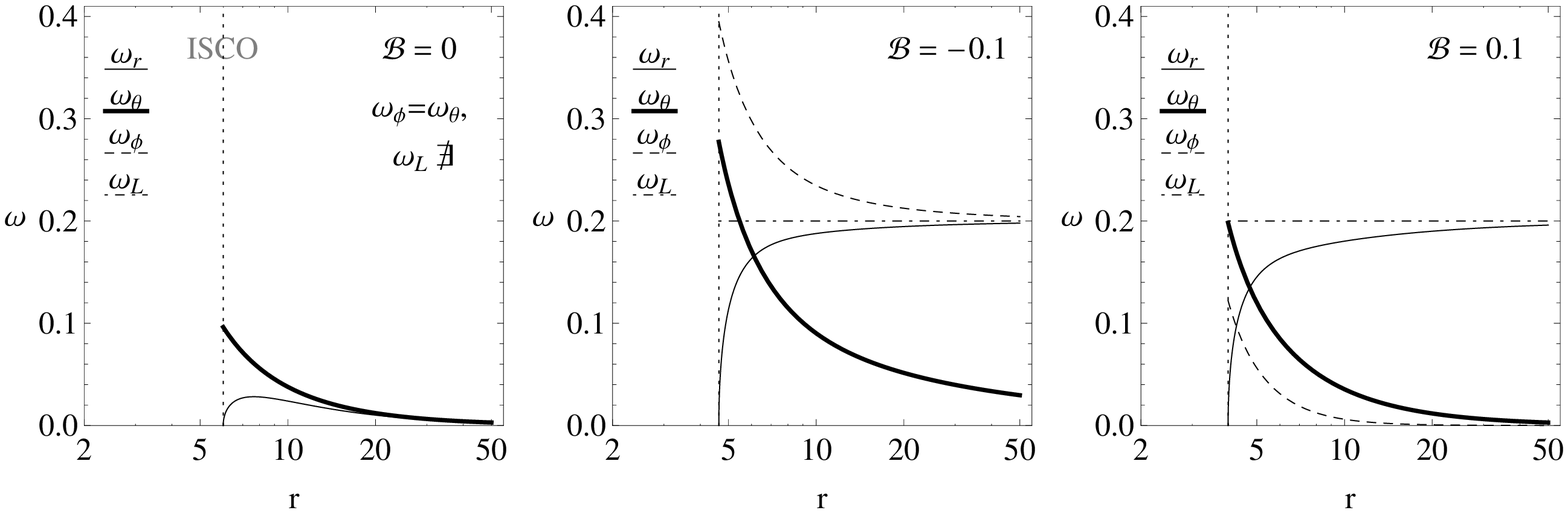}
\caption{\label{FFlocal}
Radial profiles of the locally measured fundamental frequencies $\omega_\mip$, $\omega_\mir$, $\omega_\mit$, and $\omega_\mil$ for the charged particle motion. We compare the case of uncharged particle motion with $\cb=0$ (left) to the charged particle motion with $\cb<0$ (middle) and $\cb>0$ (right). We give the radial profiles for $\cb=\pm~0.01$ and $\cb=\pm~0.1$; the same behaviour occurs for other values of the parameter $\cb$. 
}
\end{figure*}
%-------------------------------------------------------------------------%

Behaviour of the locally measured angular frequencies $\omega_\mir(r), \omega_\mit(r), \omega_\mip(r)$, and $\omega_\mil(r)$, as functions of the radial coordinate $r$, is demonstrated for significant values of the electromagnetic interaction parameter, $\cb=0$ and $\cb=\pm~0.01,0.1$, in Fig. \ref{FFlocal}. 
For small radii, $r\geq~r_{\rm ISCO}$, we see strong gravitational influence on the angular frequencies (except the Larmor frequency, of course), for large radii $r\gg~r_{\rm ISCO}$ the influence of the uniform magnetic field is prevailing.
Close to the ISCO, the latitudinal $\omega_\mit$ frequency is always larger than the radial $\omega_\mir$ frequency, $\omega_\mit>\omega_\mir$, but since $\omega_\mir~\rightarrow~\omega_\mil$ and $\omega_\mit~\rightarrow~0$ as coordinate $r$ is increasing, there always exists radius $r_{1:1}$ where $\omega_\mir=\omega_\mit$. The radial profiles of the latitudinal $\omega_\mit$ and radial $\omega_\mir$ angular frequencies do not demonstrate significant dependence on the sign of the parameter $\cb$.
The radial angular frequency $\omega_\mir$ is approaching the Larmor frequency $\omega_\mil$, from above for $\cb<0$ and from below for $\cb>0$, as the coordinate $r$ is increasing $\omega_\mir~\rightarrow~\omega_\mil$.
The Keplerian frequency $\omega_\mip$ is approaching asymptotically the Larmor frequency $\omega_\mil$ from above for $\cb<0$, but for $\cb>0$ it vanishes asymptotically.  

Behaviour of the fundamental frequencies $\omega_\mir, \omega_\mit, \omega_\mip$ and their ratios can help us to distinguish different shapes of charged particle epicyclic orbits in the vicinity of a stable circular orbit.
According to the Newtonian theory of gravitation, all the frequencies are equal, $\omega_\mir=\omega_\mit=\omega_\mip$, giving ellipse as the only possible bounded trajectory of a test particle around a gravitating spherically symmetric body. 
For uncharged particles moving around a \Schw{} black hole the relation $\omega_\mir<\omega_\mit=\omega_\mip$ holds, and there exist a periapsis shift for bounded elliptic-like trajectory implying the effect of relativistic precession that increases with decreasing radius of the orbit as the strong gravity region is entered \cite{Ste-Vie:1999:PHYSRL:}.
For charged particles orbiting a \Schw{} black hole immersed in an uniform magnetic field, the relativistic precession effect is modified by the presence of the magnetic field, and we can observe for the positive electromagnetic interaction parameters, $\cb>0$, a special family of bounded orbits with curls.  

%-------------------------------------------------------------------------%
\begin{figure*}
\includegraphics[width=\hsize]{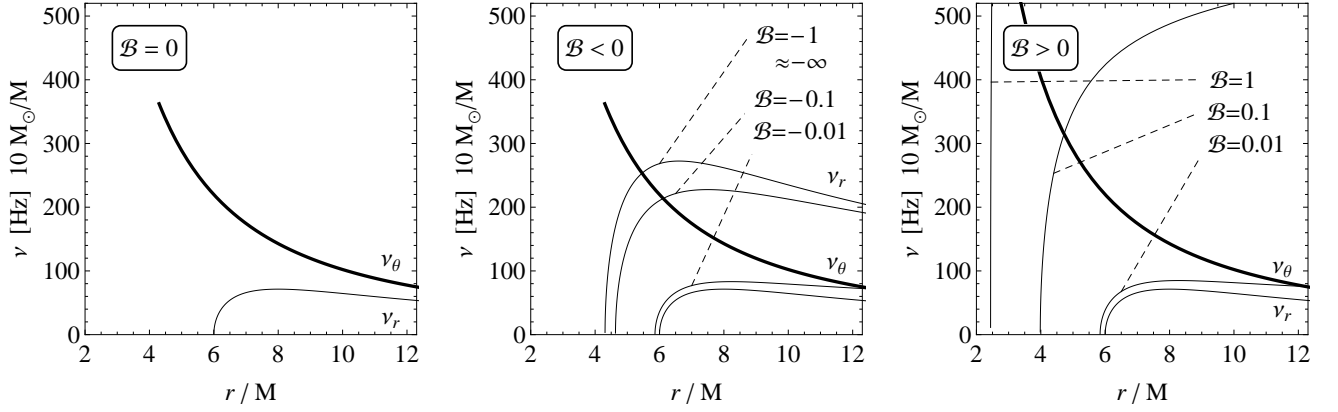}
\caption{\label{psQPO}
Radial profiles of the fundamental radial and latitudinal harmonic frequencies $\nu_\mir$ (thin) and $\nu_\mit$ (thick) related to the static distant observers. Uncharged particle frequencies (left), are compared to the charged particle frequencies for $\cb<0$ (middle), and $\cb>0$ (right). The gravitational attractor is a \Schw{} black hole with $M=10\,{M}_{\odot}$. The latitudinal frequency radial profiles, $\nu_{\mit}(r)$, are restricted to the region of existence of the stable circular orbits, limited thus by the ISCO where the radial frequency $\nu_{\mir}(r)$ vanishes. 
}
\end{figure*}
%-------------------------------------------------------------------------%

Spiral orbits which are resembling a toroidal solenoid, have to satisfy the condition $(\omega_\mir\aprx\omega_\mit)~\gg~\omega_\mip$, but this condition is not valid for charged particles orbiting a charged source in the weak gravity limit \cite{Kov:2013:EPJP:}, and we have shown that it is not possible even for the charged particles orbiting a nonrotating \Schw{} black hole placed in an uniform magnetic field. On the other hand, in the field of rotating Kerr black holes and naked singularities, the spiral orbits can exist because of the existence of relativistic orbits with low (Keplerian) angular velocity relative to distant static observers \cite{Bal-Bic-Stu:1989:BAC:,Stu:1980:BAC:}.

%%%%%%%%%%%%%%%%%%%%%%%%%%%%%%%%%%%%%%%%%%%%%%%%%%%%%%%%%%%%%%%%%%%%%%%%%%%
\subsection{Frequencies measured by distant observers}
%%%%%%%%%%%%%%%%%%%%%%%%%%%%%%%%%%%%%%%%%%%%%%%%%%%%%%%%%%%%%%%%%%%%%%%%%%%

The locally measured angular frequencies $\omega_\mir,\omega_\mit$, and $\omega_\mip$, given by
\beq
\omega_{\beta}= \frac{\d \aaa_\beta}{\d \tau},
\eeq
where $\beta \in \{\mir, \mit, \mip \} $, are connected to the angular frequencies measured by the static distant observers, $\Omega$, by the gravitational redshift transformation
\beq
 \Omega_{\beta} = \frac{\d \aaa_\beta}{\d t} = \omega_{\beta} \, \frac{\d \tau}{\d t} = \frac{\omega_{\beta}}{f(r) \ce(r)},
\eeq
where $(f(r) \ce(r) )^{-1} $ is the redshift coefficient, given by the  function $f(r)$ and the particle specific energy at the circular orbit $\ce(r)$.

If the fundamental frequencies of the small harmonic oscillations related to the distant observers, $\Omega_{\beta}$, are expressed in the physical units, their dimensionless form has to be extended by the factor $c^3/GM$. Then the frequencies of the charged particle radial and latitudinal harmonic oscillations measured by the distant observers are given by 
\beq
     \nu_{\beta} = \frac{1}{2\pi} \frac{c^3}{GM} \, \Omega_{\beta}.
\eeq

%-------------------------------------------------------------------------%
\begin{figure*}
\includegraphics[width=\hsize]{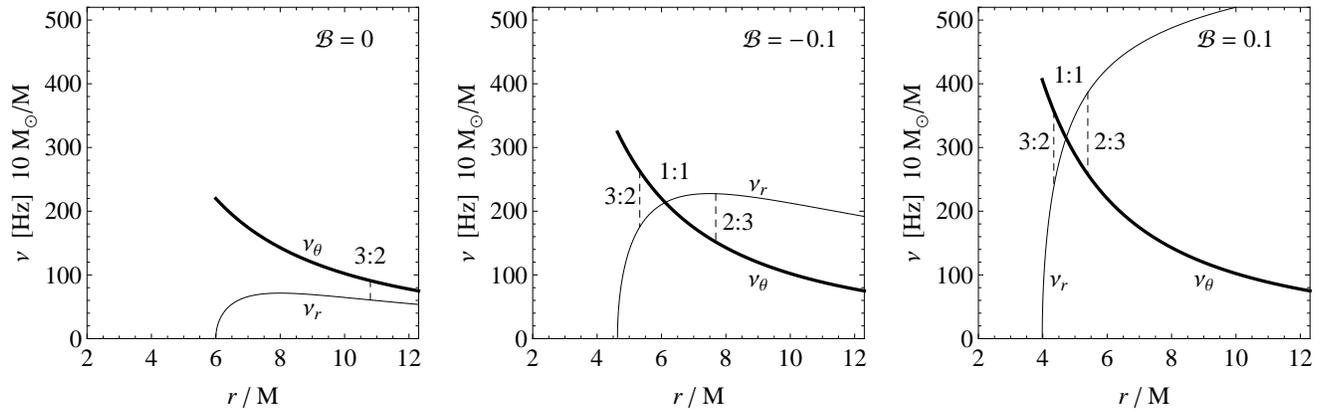}
\caption{\label{rezQPO}
Frequencies $\nu_\mit, \nu_\mir$ of small harmonic oscillations measured by static distant observers are given for the charged particles in the combined Schwarzschild and uniform magnetic fields. Position of the resonant radii are also presented.}
\end{figure*}
%-------------------------------------------------------------------------%

The dimensionless radial $\Omega_{\mir}$, latitudinal $\Omega_{\mit}$ and Keplerian $\Omega_{\mip}$ angular frequencies related to the distant observers are for the charged particle harmonic oscillations around a \Schw{} black hole immersed in an uniform magnetic field given by
\bea
\Omega^2_{\mir}(r;\cb) &=& \frac{r-6 +4\cb^2 r^3 H + 8\cb^3 (r-2) r^3 F}{r^4\left(1+4\cb^2r^2\right)}, \label{QPOsR}\\
\Omega^2_{\mit}(r;\cb) &=& \frac{1}{r^3} \label{QPOsT}, \\
\Omega^2_{\mip}(r;\cb) &=& \frac{1}{r^3 \left(2 B^2 r^2 (r-2) +2 B r F + 1 \right)}, \label{QPOsP}
\eea
where we have used the function $F(r;\cb)$ (\ref{Ffunction}) and introduced the function 
\beq
H(r;\cb) = 2 \cb^2 r^3-8 \cb ^2 r^2+8 \cb^2 r+r-4.
\eeq
In the models of the HF QPOs in the black hole systems, the epicyclic oscillatory frequencies $\nu_{\mir}, \nu_{\mit}$ are usually relevant \cite{Tor-etal:2005:ASTRA:}. Therefore, we focus our attention here on the frequencies of the radial and latitudinal harmonic oscillations. 
Frequencies $\nu_{\mir}(r), \nu_{\mit}(r)$ related to the distant observers, as functions of the radial coordinate $r$, are plotted in Figs. \ref{psQPO} and \ref{rezQPO}. 

In the pure \Schw{} black hole spacetimes ($\cb=0$), the harmonic oscillations have frequencies related to static distant observers given by expressions that are relatively very simple 
\beq
 \Omega^2_{\rm \mir (geo)}(r) = \frac{r-6}{r^4}, \quad \Omega^2_{\rm \mit (geo)}(r) = \frac{1}{r^3} .
\eeq
It is quite interesting that the latitudinal frequency of the epicyclic geodetical motion $\Omega^2_{\rm \mit (geo)}$ equals to the charged particle latitudinal frequency in the combined \Schw{} and uniform magnetic fields, see eq. (\ref{QPOsT}). This effect is caused by the direction of the Lorentz force acting on the charged particle (resulting from the external uniform magnetic field) - in the equatorial plane where the oscillations occur, it acts only in the radial direction. The motion is free, influenced only by gravitational and inertial forces in the vertical (latitudinal) direction.

%-------------------------------------------------------------------------%
\begin{figure*}
\includegraphics[width=\hsize]{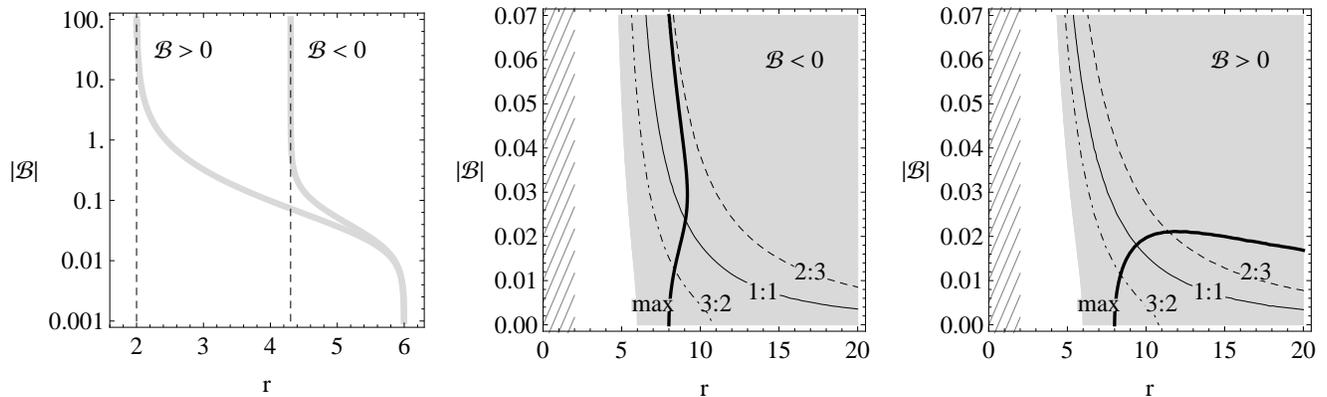}
\caption{\label{vlQPO}
Properties of the radial profile of the fundamental angular frequency of the radial harmonic oscillations, $\Omega_{\mir}(r;\cb)$, of charged particles oscillating in the combined Schwarzschild and uniform magnetic fields. 
(left) Position of the ISCO in dependence on the parameter $\cb$ that reveals from which radius the bounded motion of charged particles can start. The radial harmonic frequency vanishes at the ISCO. 
(middle and right) Regions where the particle motion is bounded and the expressions for the harmonic oscillations are relevant are greyed, region below the black hole horizon is hatched.
Solid thick curve represents the maximum of the $\Omega_{\mir}(r)$ function, various resonant radii (where $\Omega_{\mit}:\Omega_{\mir}=n:m$) are also plotted in dependence on the parameter $\cb$. 
}
\end{figure*}
%-------------------------------------------------------------------------%

%%%%%%%%%%%%%%%%%%%%%%%%%%%%%%%%%%%%%%%%%%%%%%%%%%%%%%%%%%%%%%%%%%%%%%%%%%%%%%%%%%%%%%%
\subsection{Properties of the radial profiles of frequencies of the charged particle harmonic oscillations}

We discuss properties of the radial profiles of both radial $\Omega_{\mir}$ and latitudinal (vertical) $\Omega_{\mit}$ angular frequencies of the harmonic oscillations, they are relevant in the models of the HF QPOs \cite{Tor-etal:2005:ASTRA:}. The behaviour of latitudinal harmonic angular frequency, $\Omega^2_{\mit}(r;\cb)$ (\ref{QPOsT}), is quite simple, being of the standard Newtonian character, and does not depend on the parameter $\cb$.

The zero points of the radial profile of angular frequency of the radial harmonic oscillations are given by the condition
\beq
   \Omega_{\mir}^2 (r;\cb) = 0  \label{OmNull}
\eeq
that corresponds to the limit on the existence of the stable orbits, the ISCO radii (\ref{ISCO}). Below ISCO, the particles start to collapse into black hole horizon, and no radial oscillations can occur. The local extrema of the radial profiles of angular frequency of the radial oscillations are given by the condition
\beq
    \frac{\d \Omega_{\mir}(r;\cb)}{\d r} = 0.
\eeq

The coincidence condition for the radial and latitudinal angular frequencies 
\beq
       \Omega_\mir(r;\cb) = \Omega_{\mit}(r;\cb) 
\eeq
implies the existence of only one coincidence radius $r_{1:1}$ for all values of the parameter $\cb$.

Summary of the properties of the $\Omega_\mir(r;\cb)$ function can be found in Fig. \ref{vlQPO} for both $\cb<0$ and $\cb>0$ cases. Both cases show similar behaviour for small values of the $\cb$ parameter, for larger values of the $\cb$ parameter the important radii tend to the value $r_0 \rightarrow 2$ for $\cb\rightarrow\infty$, while for $\cb\rightarrow-\infty$ we obtain the critical radius  
\beq
 r_0 \rightarrow \frac{1}{2} \left(5+\sqrt{13}\right) \doteq 4.30278,
\eeq
see Fig. \ref{vlQPO}. 

%%%%%%%%%%%%%%%%%%%%%%%%%%%%%%%%%%%%%%%%%%%%%%%%%%%%%%%%%%%%%%%%%%%%%%%%%%%
\begin{table}[!h]
\begin{center}
\begin{tabular}{c l l l}
\hline
Source & GRO 1655-40 & XTE 1550-564 & GRS 1915+105 \\
\hline \hline
$ \nu_{\rm U}$ [Hz] & $447${\lin}$453$ & $273${\lin}$279$ & $165${\lin}$171$ \\
$ \nu_{\rm L}$ [Hz] & $295${\lin}$305$ & $179${\lin}$189$ & $108${\lin}$118$ \\
$ M/{M}_\odot $ & $6.03${\lin}$6.57$ & $8.5${\lin}$9.7$ & $9.6${\lin}$18.4$ \\
$ a $ & $0.65${\lin}$0.75$ & $0.29${\lin}$0.52$ & $0.98${\lin}$1$ \\
\hline
\end{tabular}
\caption{Observed twin HF QPO data for the three microquasars, and the restrictions on mass and spin of the black holes located in them, based on measurements independent of the HF QPO measurements given by the spectral continuum fitting \cite {Sha-etal:2006:ApJ:,Rem-McCli:2006:ARAA:}.} \label{tab1}
\end{center}
\end{table}
%%%%%%%%%%%%%%%%%%%%%%%%%%%%%%%%%%%%%%%%%%%%%%%%%%%%%%%%%%%%%%%%%%%%%%%%%%%

%%%%%%%%%%%%%%%%%%%%%%%%%%%%%%%%%%%%%%%%%%%%%%%%%%%%%%%%%%%%%%%%%%%%%%%%%%%%%%%%%%%%%%%
\section{Charged particle oscillations at resonance 3:2 and 2:3 radii modelling the twin HF QPOs in the microquasars \label{observations}}
%%%%%%%%%%%%%%%%%%%%%%%%%%%%%%%%%%%%%%%%%%%%%%%%%%%%%%%%%%%%%%%%%%%%%%%%%%%%%%%%%%%%%%%

The quasi-harmonic character of the motion of charged particles trapped in a toroidal space around the equatorial plane of magnetized black holes suggests an interesting astrophysical application related to the HF QPOs observed in the LMXB systems containing a black hole or a neutron star, or in an active galactic nuclei. Some of the~HF~QPOs come in pairs of the~upper and lower frequencies ($\nu_{\mathrm{U}}$, $\nu_{\mathrm{L}}$) of twin peaks in the~Fourier power spectra. Since the~peaks of high frequencies are close to the~orbital frequency of the~marginally stable circular orbit representing the~inner edge of Keplerian discs orbiting black holes (or neutron~stars), the~strong gravity effects must be relevant in explaining HF~QPOs \cite{Tor-etal:2005:ASTRA:}.

%%%%%%%%%%%%%%%%%%%%%%%%%%%%%%%%%%%%%%%%%%%%%%%%%%%%%%%%%%%%%%%%%%%%%%%%%%%
\begin{center}
\begin{table*}[ht]
{\small
\hfill{}
\begin{tabular}{|c c @{\quad} l @{\quad} | l @{\quad} l @{\quad} l @{\quad} |} 
\hline
& & $\cb$ & $B_{\rm e-}$ [mGs] & $B_{\rm p+}$ [Gs] & $B_{\rm Fe}$ [Gs] \\
\hline \hline
$-$ & 3:2 & 0.063{\lin}0.110 & 0.17{\lin}0.26 & 0.31{\lin}0.47 & 17.38{\lin}26.60 \\
$-$ & 2:3 & 0.113{\lin}0.370 & 0.31{\lin}0.88 & 0.56{\lin}1.61 & 31.18{\lin}89.47 \\
$+$ & 3:2 & 0.041{\lin}0.056 & 0.11{\lin}0.13 & 0.20{\lin}0.24 & 11.31{\lin}13.54 \\
$+$ & 2:3 & 0.048{\lin}0.058 & 0.13{\lin}0.14 & 0.24{\lin}0.25 & 13.25{\lin}14.03 \\
\hline
\end{tabular}
}
\hfill{}
\caption{Boundary values of the magnetic field parameter $\cb$ estimated for the microquasar XTE~1550-564. 
First part of this table contains restrictions on the $\cb$ parameter for slowly rotating source XTE~1550-564 given by the resonance oscillation model of twin HF OPOs, see Fig \ref{QPOfit}.
The mass of XTE~1550-564 source implies limits on the magnetic field strength. We present such limits for electron $B_{\rm e-}$, proton $B_{\rm p+}$ and partially ionized (one electron lost) iron atom $B_{\rm Fe-}$ in Gauss units in second part.
\label{tab2}
} 
\end{table*}
\end{center}
%%%%%%%%%%%%%%%%%%%%%%%%%%%%%%%%%%%%%%%%%%%%%%%%%%%%%%%%%%%%%%%%%%%%%%%%%%%

%%%%%%%%%%%%%%%%%%%%%%%%%%%%%%%%%%%%%%%%%%%%%%%%%%%%%%%%%%%%%%%%%%%%%%%%%%%
\begin{center}
\begin{table*}[ht]
{\small
\hfill{}
\begin{tabular}{|c c @{\quad} l @{\quad} | l @{\quad} l @{\quad} l @{\quad} |} 
\hline
& & $\cb$ & $Q\times 10^{6}$ [esu] & $m\times 10^{-16}$ [g] & $n\times 10^8$ \\
\hline \hline
$-$ & 3:2 & 0.063{\lin}0.110 & 0.90 {\lin} 1.38 &  5.33 {\lin} 3.49 &  3.18 {\lin} 2.08 \\
$-$ & 2:3 & 0.113{\lin}0.370 & 1.61 {\lin} 4.63 &  2.97 {\lin} 1.04 & 1.78 {\lin} 6.19 \\
$+$ & 3:2 & 0.041{\lin}0.056 & 0.59 {\lin} 0.70  &  8.19 {\lin} 6.85 & 4.89 {\lin} 4.09 \\
$+$ & 2:3 & 0.048{\lin}0.058 & 0.69 {\lin} 0.73 & 7.00 {\lin} 6.61 & 4.18 {\lin} 3.95 \\
\hline
\end{tabular}
}
\hfill{}
\caption{Boundary values of the magnetic field parameter $\cb$ estimated for the microquasar XTE~1550-564. 
Last three columns represent the inverse estimations of the specific charge $Q = q/m$, mass $m$ and number of particles $n$ of the oscillating test object (or blob), for the fixed value of the strength of the magnetic field $B=10^8$ Gs. We assume the one-electron ionization, in the other words the blob has to be slightly ionized. More stronger ionization increases the mass $m$ and the number of particles in the blob for the corresponding value defining by the number of electrons removed away from initially neutral blob. The blob mass is expressed in units of the mass of the neutron $n$ in the last column.
\label{tab3}
} 
\end{table*}
\end{center}
%%%%%%%%%%%%%%%%%%%%%%%%%%%%%%%%%%%%%%%%%%%%%%%%%%%%%%%%%%%%%%%%%%%%%%%%%%%

%-------------------------------------------------------------------------%
\begin{figure*}
\includegraphics[width=\hsize]{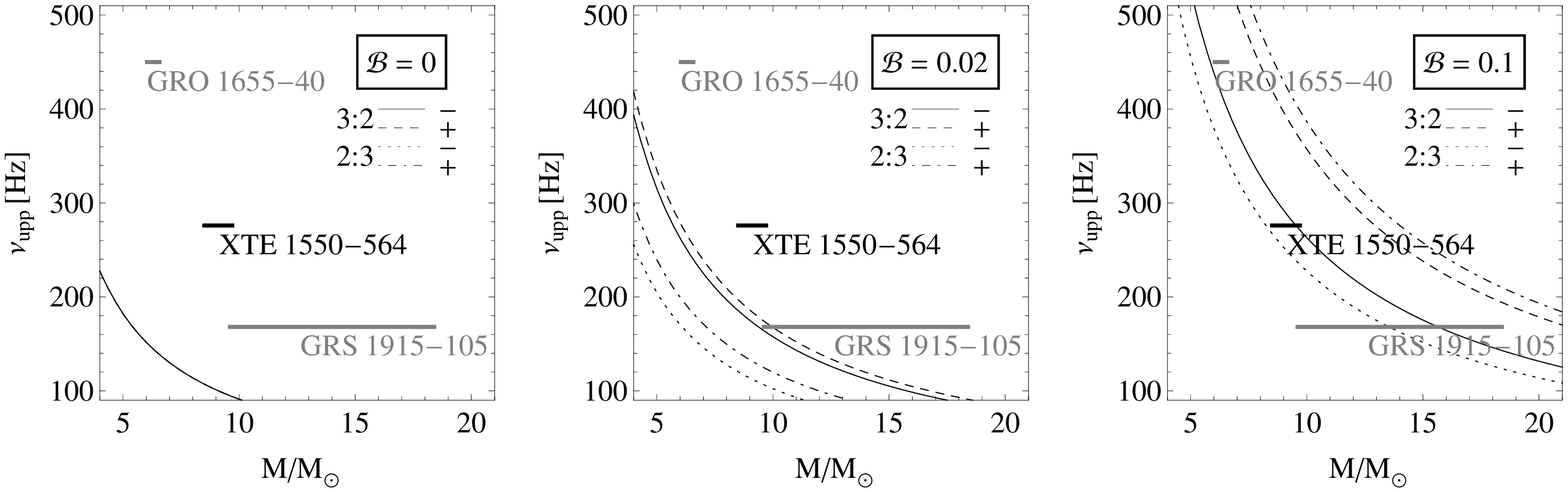}
\caption{\label{QPOfit}
The upper oscillation frequency $\nu_{\rm U}$ at the 3:2 or 2:3 resonance radii, calculated for both the negative $\cb<0$ ($-$) and positive $\cb>0$ ($+$) values of the magnetic parameter, and compared to the mass-limits obtained from observations of the slowly rotating microquasars XTE 1550-564. The other two, fast rotating, microquasars GRO 1655-40 and GRG 1915-105 are also presented.
}
\end{figure*}
%-------------------------------------------------------------------------%

Before the~twin peak HF~QPOs have been discovered in microquasars, first by \cite{Str:2001:ASTRJ2L:}, and the~3\,:\,2 ratio has been pointed out, \cite{Klu-Abr:2000:ASTROPH:} suggested that these QPOs should have rational ratios, because of the~resonances in oscillations of nearly Keplerian accretion disks; see also \cite{Ali-Gal:1981:GRG:}. It seems that the~resonance hypothesis is now well supported by observations, and that the~3\,:\,2 ratio ($2\nu_{\mathrm{U}} = 3\nu_{\mathrm{L}}$) is seen most often in the twin HF QPOs in the~LMXB containing black holes (microquasars). Unfortunately, neither of the recently discussed models based on the frequencies of the harmonic geodesic epicyclic motion (based on Keplerian discs or slender tori \cite{Tor-etal:2005:ASTRA:,Rez-etal:2003:MNRAS:}) is able to explain the HF QPOs in all the three microquasars assuming their central attractor to be a black hole \cite{Tor-etal:2011:ASTRA:}. 

We can assume applicability of the parametric resonance, discussed in \cite{Lan-Lif:1969:Mech:}, focusing attention to the case of the frequency ratios $\nu_{\mit}:\nu_{\mir} = 3:2$ or $\nu_{\mit}:\nu_{\mir} = 2:3$, as the observed values of the twin HF QPO frequencies for GRO 1655-40, XTE 1550-564 and GRS 1915+105 sources show clear ratio
\beq
 \nu_{\rm U} : \nu_{\rm L} = 3 : 2
\eeq
for the upper $\nu_{\rm U}$ and lower $\nu_{\rm L}$ frequencies, see Tab. \ref{tab1}. We identify directly the frequencies $\nu_{\rm U}, \nu_{\rm L}$ with $\nu_{\mit}, \nu_{\mir}$ or $\nu_{\mir}, \nu_{\mit}$  frequencies. In contrast to the standard resonance epicyclic model (without external magnetic field), the oscillating charged particles in the combined \Schw{} and uniform magnetic fields allow both frequency ratios
\beq
  \nu_{\mit} : \nu_{\mir} = 3 : 2, \quad \nu_{\mit} : \nu_{\mir} = 2 : 3.
\eeq
Since $ r_{3:2} < r_{2:3} $ (see Figs. \ref{rezQPO} and \ref{vlQPO}), we call the first resonance radius, where ${\nu_\mit:\nu_{\mir}=3:2}$, the inner one, and the second resonance radius, where ${\nu_\mit:\nu_{\mir}=2:3}$, the outer one. Note that for the oscillating charged particles also the ${1:1}, {1:2}, {2:1}$, or other, resonant frequency ratios can enter the play - these can be relevant in other sources where such frequency ratios are observed, see, e.g., \cite{Lach-Cze-Abr:2013:arxiv:}, or for the other twin frequencies observed in the microquasar GRS~1915+105.

The procedure of fitting the charged particle oscillation frequencies to the observed frequencies is presented in Fig. \ref{QPOfit}, for all the three microquasars GRS~1915+105, XTE~1550-564, and GRO~1655-40. From the restrictions on the spacetime mass parameter $M$ for each of the sources, see Tab. \ref{tab1}., we  obtain simultaneously restrictions on the external uniform magnetic field through restrictions on the magnetic parameter $\cb$. So far we have not considered the sources rotation that can modify the radial profiles of the vertical and horizontal frequencies -- the effect of the black hole rotation is going to be studied in a future work, but our preliminary results show that the fitting can be done even for the extremely fast rotating microquasar GRS~1915+105. 

We use the fitting procedure in the case of relatively slowly ($a\aprx0.4$) rotating XTE~1550-564 source as an useful example. Detailed results for the XTE~1550-564 source are given in Tab. \ref{tab2} and \ref{tab3}, where we  present typical values of the magnetic field related to astrophysically realistic situations. Strong magnetic fields are not necessary - for electrons, the required magnetic field strength $B_{\rm e-}\aprx0.1$\,mGs is comparable to the magnetic field strength in the heliosphere, for protons, $B_{\rm p+}\aprx0.2$\,Gs it is comparable to Earth's magnetic field at its surface, and for partially ionized (one electron lost) iron atom $B_{\rm Fe}\aprx10$\,Gs is comparable to the magnetic field strength in Earth's core \cite{Zel-Ruz-Sok:1983:MagAstro:}.

%%%%%%%%%%%%%%%%%%%%%%%%%%%%%%%%%%%%%%%%%%%%%%%%%%%%%%%%%%%%%%%%%%%%%%%%%%%
\section{Conclusions}
%%%%%%%%%%%%%%%%%%%%%%%%%%%%%%%%%%%%%%%%%%%%%%%%%%%%%%%%%%%%%%%%%%%%%%%%%%%

External uniform magnetic field $\cb$ can strongly influence the charged particle motion around a magnetized black hole. In dependence on initial inclination angle $\theta_0$, the charged particle can escape to infinity along the $z$ axis related to the direction of the magnetic field, it can start to move chaotically, or it can start to oscillate in a harmonic or quasi-harmonic regime. Even small magnetic field, $\aprx10$\,Gs, can significantly influence the charged particle trajectory, if the specific charge of test particle is large enough. The oscillatory motion can be curled, giving a clear resemblance to the Larmor precession in a pure magnetic field. 

We have shown that assuming relevance of the resonant phenomena of the radial and vertical (latitudinal) oscillations at their frequency ratio $3:2$, the oscillatory frequencies of charged particle in uniform magnetic field of a magnetized black hole can be well related to the frequencies of the twin $3:2$ HF QPOs observed in the microquasars GRS~1915+105, XTE~1550-564, GRO~1655-40. We can conclude that simple radial and latitudinal charged particle quasi-harmonic oscillations can be considered as one of the possible explanations of the HF QPOs occurring in the field of compact objects. 

The results of the procedure of fitting the frequencies of the twin HF QPOs in the three microquasars for the charged particle oscillations are similar to those obtained in \cite{Stu-Kol:2014:PHYSR4:} in the framework of the string loop oscillation model \cite{Kol-Stu:2013:PHYSR4:}, indicating some similarities in the dynamics of charged particles and string loops.

The effect of even weak external uniform magnetic field of a magnetized Schwarzschild black hole on the radial and latitudinal frequencies of harmonic oscillations around a stable equatorial orbit is substantial; the magnetic field influence can erase all ``strange'' effects, like those coming from the external dimension in the braneworld model \cite{Kot-Stu-Tor:2008:CLAQG:,Stu-Kot:2009:GRG:}. 

The single isolated charged particle dynamics can be imprinted in description of diluted collisionless plasma toroidal structures \cite{Cre-etal:2013:ApJS:}, the kinetic description of collisionless plasma can effectively reflect the epicyclic motion of charged particles \cite{Cre-Stu:2013:PHYSRE:}. Full magnetohydrodynamics simulations at the present state are not able to reflect the high-frequency QPOs, only low-frequency QPOs have been reflected in some models \cite{Mac-Mat:2008:PAJC:}. The charged particle epicyclic motion can still be used for dynamics of ionized blob structures created by instabilities or by irradiation in otherwise neutral accretion disk \cite{Stu-Kot-Tor:2013:ASTRA:}.

We can conclude that the charged particle quasi-harmonic oscillation model has to be considered seriously, but deeper details have to be studied and the results have to be confronted with the results of modelling other signatures of the strong gravity related to the optical phenomena, such as the spectral line profiles or light curves. 

%%%%%%%%%%%%%%%%%%%%%%%%%%%%%%%%%%%%%%%%%%%%%%%%%%%%%%%%%%%%%%%%%%%%%%%%%%%
\section*{Acknowledgments}
%%%%%%%%%%%%%%%%%%%%%%%%%%%%%%%%%%%%%%%%%%%%%%%%%%%%%%%%%%%%%%%%%%%%%%%%%%%

The authors would like to express their acknowledgments for the Institutional support of the Centre for Theoretical Physics and Astrophysics at the Faculty of Philosophy and Science of the Silesian University in Opava.
The authors acknowledge the Albert Einstein Centre for Gravitation and Astrophysics supported by the Czech Science Foundation grant No.~14-37086G and the internal student grant of the Silesian University SGS/23/2013.

%%%%%%%%%%%%%%%%%%%%%%%%%%%%%%%%%%%%%%%%%%%%%%%%%%%%%%%%%%%%%%%%%%%%%%%%%%%
%%%%%%%%%%%%%%%%%%%%%%%%%%%%%%%%%%%%%%%%%%%%%%%%%%%%%%%%%%%%%%%%%%%%%%%%%%%

%\section*{References}

%\bibliographystyle{iopart-num}
%\input{C:/DOC/WORK/TXT/reference/refdef}
%\bibliography{C:/DOC/WORK/TXT/reference/reference}

\def\prc{Phys. Rev. C}
\def\pre{Phys. Rev. E}
\def\prd{Phys. Rev. D}
\def\jcap{Journal of Cosmology and Astroparticle Physics}
\def\mnras{Mon. Not. R. Astron Soc.}
\def\apj{The Astrophysical Journal}
\def\aap{Astron. Astrophys.}
\def\pasj{Publications of the Astronomical Society of Japan}
\def\pasa{Publications Astronomical Society of Australia}
\def\nat{Nature}
\def\physrep{Phys. Rep.}
\def\apjs{The Astrophysical Journal Supplement}
\def\apjl{Astrophysical Journal Letters}
\def\araa{Annual Review of Astronomy and Astrophysics}

\end{document}